\documentclass[a4paper,fleqn,usenatbib]{mnras}
\usepackage{upgreek}
\usepackage{graphicx}   
\usepackage{amsmath}    
\usepackage{amssymb}    
\usepackage[export]{adjustbox}
\usepackage{slashbox}

\renewcommand{\div}{\ensuremath{-}}

\DeclareRobustCommand{\uppartial}{\text{\rotatebox[origin=t]{20}{\scalebox{0.95}[1]{$\partial$}}}\hspace{-1pt}}

\begin{document}
\title[Super-Eddington accretion onto a magnetized neutron star]{Super-Eddington accretion onto a magnetized neutron star}
\author[A. Chashkina et al.]{Anna Chashkina,$^{1,2,3,4}$\thanks{E-mail:
		anna.chashkina@utu.fi}  Pavel Abolmasov$^{1,2,4}$ and Juri Poutanen$^{1,3,4}$ \\
	$^{1}$Tuorla Observatory, Department of Physics and Astronomy, University of Turku, V\"ais\"al\"antie 20, FI-21500 Piikki\"o, Finland\\
	$^{2}$Sternberg Astronomical Institute, Moscow State University, Universitetsky pr. 13, Moscow 119992, Russia\\
	$^{3}$Nordita, KTH Royal Institute of Technology and Stockholm University, Roslagstullsbacken 23, SE-10691 Stockholm, Sweden\\
    $^{4}$Kavli Institute for Theoretical Physics, University of California, Santa Barbara, CA 93106, USA 
}

\date{Accepted ---. Received ---; in
  original form --- }

\pagerange{\pageref{firstpage}--\pageref{lastpage}} \pubyear{2017}

\maketitle

\label{firstpage}

\begin{abstract}
	 Most of ultraluminous X-ray sources are thought to be objects accreting above their Eddington limits. 
    In the recently identified class of ultraluminous X-ray pulsars, accretor is a neutron star and thus has a fairly small mass with a small Eddington limit. 
		The accretion disc structure around such an object affects important observables such as equilibrium period, period derivative and the size of the magnetosphere. 
		We propose a model of a nearly-standard accretion disc interacting with the magnetosphere only in a thin layer near the inner disc rim. 
		Our calculations show that the size of the magnetosphere may be represented as the classical Alfv\'en radius times a dimensionless factor $\xi$ which depends on the disc thickness only. 
		In the case of radiation-pressure-dominated disc, the size of the magnetosphere does not depend on the mass accretion rate. 
		In general, increasing the disc thickness leads to a larger magnetosphere size in units of the Alfv\'en radius. 
		For large enough mass accretion rates and magnetic moments, it is important to take into account not only the pressure of the magnetic field and the radiation pressure inside the disc, but also the pressure of the radiation produced close to the surface of the neutron star in accretion column. 
		The magnetospheric size may increase by up to factor of two as a result of  the effects related to the disc thickness and the irradiation from the central source. 
		Accounting for these effects  reduces the estimate of the neutron star magnetic moment by a factor of several.
		
\end{abstract}

\begin{keywords} 
accretion, accretion discs -- hydrodynamics -- magnetic fields -- stars: neutron -- X-rays: binaries
\end{keywords}

\section{Introduction}

Magnetospheric accretion is a very important topic in astrophysics, especially for magnetized neutron stars. 
Interaction of falling matter with magnetosphere determines observational properties of neutron stars such as their equilibrium period and the period derivative $\dot p$.  
At low mass accretion rates, matter enters the magnetosphere through a quasi-spherical envelope  \citep{SPH}. 
When the angular momentum of the matter is large enough, accretion disc is likely to be formed. 
The disc forms when the angular momentum is larger than the Keplerian angular momentum at the magnetospheric radius $R_{\rm in}$ \citep{IS75}: 
\begin{equation}
\zeta \Omega R^2_{\rm G}>\sqrt{GMR_{\rm in}},
\end{equation}
where $\zeta$ is a numerical factor that depends on the structure of the wind, $R_{\rm G}$ is the Bondi radius, $M$ is the neutron star mass.
In the disc case, the neutron star luminosity is determined by the mass loss of the star. 
For the case of accretion via wind, the mass accretion rate  is lower than the mass loss of the companion star by a factor of $(R_G/a)^2\ll 1$, where $a$ is the binary separation. 
The Roche-lobe overflow in high-mass X-ray binaries (HMXB) is a relatively rare phenomenon (the only persistent Galactic source of this kind is probably SS~433, see \citealt{Fabrika04}).  
HMXB with Roche-lobe overflow have higher luminosities due to a larger accretion rate and can be easily observed in other galaxies. 

There are two approaches to magnetospheric accretion.
The first one considers diamagnetic, finite-conduction disc threaded by stellar magnetic fields over a wide range of radii. 
The pioneers of this approach were \citet{GL77}; other works based on this idea by \citet{KR07} and \citet{Wang87} used slightly modified magnetic field distribution in the disc as well as non-Keplerianity near the inner edge of the disc. 
The second approach is to consider a disc interacting with the stellar magnetic field only in a very narrow region near the inner boundary of the disc (see, for example, recent simulations by \citealt{Parfrey16,PSB15}). 
One of the first works by \citet{Scharlemann78} assumes diamagnetic currents that are modulated by a single ring current located at the edge of the magnetosphere. 
Further developing this approach \citet{Aly80} studied the influence of non-radial magnetic forces. 
The instabilities due to velocity differences at the disc-magnetosphere boundary were studied by \citet{Anzer80}. 
Another effect of the angular velocity difference is opening of some magnetic lines considered by \citet{Aly90}, that can produce outflows along the open field lines \citep{Shu94,LRBK95}.  

Infalling matter creates a hot polar cap or an accretion column \citep{BS76} that radiates X-ray emission at several keV.
The structure of the accretion column, however, depends on how the matter enters the magnetosphere and, specifically, upon the radius of the magnetosphere $R_{\rm in}$.
Magnetospheric radius can be  estimated  from equality of the disc ram pressure and the magnetic pressure:
\begin{equation}\label{eq:alfven}
R_{\rm in}=\xi R_{\rm A} = \xi \left(\displaystyle\frac{\mu^2}{2\dot M \sqrt{2GM}} \right)^{2/7},
\end{equation}
where $R_{\rm A}$ is known as Alfv\'{e}n, or Alfvenic, radius, $\mu$ is the neutron star magnetic moment and $\dot M$ is the accretion rate. 
The dimensionless coefficient $\xi$ is determined by the accretion flow structure. 
For instance, for spherical accretion, $\xi=1$. 
Analytical models of disc accretion onto magnetized neutron star predict $\xi$ from $\simeq 0.5$ \citep{GL77,KR07,K91} to $1$ \citep{Wang96}. 
MHD simulations for small magnetospheres of young stars and white dwarfs  give $\xi=0.4-0.5$ \citep{Long05,Bessolaz08}.  
Recent simulations by \citet{romanova2013} show variable $\xi\simeq 0.77 ({\mu^2}/{\dot M})^{-0.086}$ that may be understood as different scaling of the magnetosphere size with the basic parameter set, roughly as $R_{\rm in} \propto (\mu^2/\dot{M})^{1/5}$. 
This is similar to the predictions of some analytical models like that of  \citet{Spruit90}. 

Until recently, the discs in all the X-ray pulsars were well understood as thin, gas-pressure dominated, and nearly Keplerian, and trimmed from the inside by the neutron star magnetosphere. 
Even for high mass accretion rates the size of the magnetosphere is usually large enough for the accretion disc to remain very far from the local Eddington limit. 
The   accretion disc thickness at high enough mass accretion rates is determined by radiation pressure (the so-called zone A, see \citealt{SS73}) and equals to 
\begin{equation}
H \simeq \dfrac{3}{8\uppi} \dfrac{\kappa \dot{M}}{c},
\end{equation}
where $\kappa=0.34$~cm$^{2}$g$^{-1}$ is the (Thomson scattering) opacity. 
At some radius known as spherization radius \citep{SS73}, the disc thickness becomes larger than the disc radius.
Even if the Eddington luminosity $L_{\rm Edd}={4\uppi GMc}/{\kappa}$ is exceeded by a considerable amount, energy release in the accretion disc is likely to remain insufficient for super-Eddington accretion unless spherization radius becomes larger than the radius of the magnetosphere. 
For this to take place, the total luminosity must exceed the Eddington value by a factor of about $R_{\rm in}/ R_*$ (where $R_*$ is the neutron star radius) that is of the order of hundreds or thousands for X-ray pulsars. 
Therefore, here we will concentrate on the quite likely case when the total luminosity is above the Eddington limit but the disc is everywhere sub-critical. 

 Column luminosity is not strictly limited by the classical Eddington limit because of geometrical and opacity effects \citep{BS76,MST15}. It is quite possible for a neutron star to exceed the Eddington limit by a factor of $ \sim 10^2-10^3$.
Such objects can shine very bright in the X-ray range and explain at least some part of the population of ultraluminous X-ray sources (ULXs) in nearby galaxies.  

Recently several ULXs have been shown to have coherent pulsations at periods $\sim 1$~s \citep{Bach14,2017Sci...355.817I,Israel17}, that identifies these objects with neutron stars rather than black holes. 
The high luminosities  ($\sim 10^{40}-10^{41}{\rm erg\,s^{-1}}$) of these sources is well in excess of the Eddington limit for a neutron star that clearly points to the importance of radiation pressure leading to a high disc thickness.
Most of the radiation in super-Eddington pulsars is released in the accretion column. 
The accretion disc, at the same time, becomes illuminated by the radiation flux exceeding the local Eddington limit, and thus its structure as well as the position of the disc-magnetosphere interface should be affected significantly.  
Both radiation from the central source (accretion column) and the magnetic field of the neutron star create a radial pressure gradient within the disc, violating one of the basic assumptions of the standard disc theory. 

In this paper we develop a model of the accretion disc with deviations from the Keplerian rotation caused by the disc thickness and by the influence of magnetic field pressure at its inner edge. 
We assume that the disc-magnetosphere interaction is confined to a thin layer at the inner boundary of the disc. 
We also study the influence of radiation pressure from accretion column on the disc structure and the magnetospheric radius for ULX-pulsars. 

In Section \ref{sec:model} we describe our model.
Techniques used to solve the equations are given in Section~\ref{sec:solve}. 
Our main results are presented in Section~\ref{sec:results} and discussed in Section~\ref{sec:discuss}. 

\section{The model}\label{sec:model}

In this section we give the whole system of equations that describes the structure of the disc. 
We start from the equations of hydrodynamics. 
The momentum equation in the vector notation takes the form: 
\begin{equation}\label{eq:eulervec}
\displaystyle\frac{ \uppartial \mathbfit{v}}{\uppartial t}+(\mathbfit{v}\cdot \nabla)\mathbfit{v}=-\displaystyle\frac{1}{\rho}\nabla P-\nabla \varphi+\mathbfit{N},
\end{equation}
where $\rho$ is density, $P$ is pressure, and $\mathbfit{v}$ is the velocity field in the disc and $\varphi$ is gravitational potential.
The last viscous term $\mathbfit{N}$ does not affect the radial and vertical components of the Euler equation. 
The azimuthal component of the equation is equivalent to the equation of angular momentum transfer that will be considered separately.
In steady state and axial symmetry, the radial component of this equation in cylindrical coordinate system can be rewritten as:
\begin{equation}\label{eq:eulerrad}
v_{R}\displaystyle\frac{\uppartial v_{R}}{\uppartial R}+v_{z}\displaystyle\frac{\uppartial v_{R}}{\uppartial z}-\Omega^2 R
=-\displaystyle\frac{1}{\rho}\frac{\uppartial P}{\uppartial R}-\displaystyle\frac{GM}{R^2}.
\end{equation}

Here $P$ is the pressure, $v_R$ is the radial and $v_z$ is the vertical velocity and $\Omega$ is the angular velocity in the disc. 
Radiation pressure of the central source shifts the radial force equilibrium of the Keplerian disc ($\Omega^2R=GM/R^2$), and the radial pressure gradient can no more be neglected  in the inner parts of the disc. 
This leads to two effects: the angular velocity of the disc becomes non-Keplerian, 
and the inner parts of the disc are additionally decelerated in radial direction. 
Thus in this case, the radial velocity is even smaller than in the standard disc, allowing us to neglect the first term, quadratic in $v_R$, on the left-hand side of equation~(\ref{eq:eulerrad}). 
Because we expect $v_{z}$ to be also small,  the second term can be neglected as well.
In general, in thin accretion discs, first two terms in equation ~(\ref{eq:eulerrad}) are of the order \citep[see e.g.][]{urpin84}

\begin{equation} 
\displaystyle\frac{v_{R, z}^2}{R} \sim \left(\alpha \left(\displaystyle\frac{H}{R}\right)^2 \right)^2 \Omega_{\rm K}^2 R \ll \displaystyle\frac{1}{\rho} \displaystyle\frac{{\rm d}p}{{\rm d}r} \sim \left(\displaystyle\frac{H}{R}\right)^2 \Omega_{\rm K}^2 R.
\end{equation}
Integrating equation~(\ref{eq:eulerrad}) over the vertical coordinate $z$, or, more precisely, over $\rho {\rm d}z$, yields:
\begin{equation}\label{eq:omega}
\Omega^2 R=\displaystyle\frac{1}{\Sigma}\frac{\uppartial \Pi}{\uppartial R}+\displaystyle\frac{GM}{R^2}.
\end{equation}
Here  $\Pi=\int^{H}_{-H}{p{\rm d}z}$ is the vertically integrated pressure, $\Sigma=\int^{H}_{-H}\rho {\rm d}z$ is the surface density, and  $H$ is the thickness of the disc. 

To relate the vertically-integrated and equatorial-plane parameters of the disc, it is important to make assumptions about the vertical structure of the disc, which is discussed in detail in Appendix \ref{sec:appendix}. 
From hydrostatic equilibrium together with the assumptions about the vertical structure, we have the central pressure in the disc:
\begin{equation}\label{eq:eulervert}
P_{\rm c}=\rho_{\rm c} \dfrac{GM}{R^3}\dfrac{H^2}{4},
\end{equation}
where $\rho_{\rm c}$ is the central (mid-plane) density.
The vertical structure together with equation (\ref{eq:eulervert}) provide us with the thickness of the disc:
\begin{equation}\label{eq:thickness}
H=\displaystyle\sqrt{\frac{5\Pi}{\Sigma}\frac{R^3}{GM}}.
\end{equation}
Following the standard accretion disc theory we assume alpha-prescription \citep{SS73}:
\begin{equation}\label{eq:alphapr}
W_{r\phi}=\alpha \Pi,
\end{equation}
where $W_{r\phi}$ is the $r\phi$-component of the vertically integrated viscous stress tensor. 
Angular momentum conservation equation is:
\begin{equation}\label{eq:momconst}
\dot M\displaystyle\frac{{\rm d}(\Omega R^2)}{{\rm d}R}=\displaystyle\frac{{\rm d}}{{\rm d}R}(2\uppi R^2 W_{r\phi}). 
\end{equation}
We assume $\dot M$ constant with radius that allows straightforward integration of this equation.
Following \citet{SS73}, we assume that the energy released in the accretion disc is radiated locally from its surface. 
In this case, the local energy release will be coupled to the radiation emitted from the disc surface  and to the angular velocity gradient in the disc  allowing to link the rotation law of the disc $\omega(R)$ with the vertically-integrated pressure. 
The pressure is itself linked to the energy release rate by the vertical radiation diffusion equation
\begin{equation}\label{eq:diffuse}
F=-D\nabla_z \epsilon = - D \dfrac{{\rm d}\epsilon}{{\rm d}z},
\end{equation}
where $F$ is the vertical energy flux,  $\epsilon=aT^4$ is the radiation energy density and

\begin{equation}\label{eq:dif}
 D=c/(3\kappa \rho)
 \end{equation}
  is the diffusion coefficient.
Local energy dissipation rate is 
\begin{equation}\label{eq:Edissip}
\displaystyle\frac{{\rm d}F}{{\rm d}z}=\alpha PR\displaystyle\left| \frac{{\rm d}\Omega}{{\rm d}R}\right|,
\end{equation}
 where $P$ is the total (gas+radiation) pressure:
\begin{equation}
P=nkT+\displaystyle\frac{aT^4}{3},
\end{equation}
where $n\simeq \dfrac{2\rho}{m_{\rm p}}$ is the proton or electron number density for proton-electron plasma and $a$ is the radiation constant.
Solving equations (\ref{eq:diffuse}) and  (\ref{eq:Edissip}) one obtains the following relation between the surface temperature $T_{\rm s}$ and the central temperature $T_{\rm c}$ :
\begin{equation}\label{eq:surft}
T^4_{\rm s}=T_{\rm c}^4-\displaystyle\frac{73}{120}\displaystyle\frac{1}{a}\displaystyle\frac{\alpha \kappa \rho_{\rm c} P_{\rm c} RH^2}{c},
\end{equation}
where the dimensionless coefficient is derived in Appendix \ref{sec:appendix}. 
Assuming that the surface temperature $T_{\rm s}$ equals the effective temperature determined by the overall energy release in the disc, 
\begin{equation} 
2\sigma_{\rm SB} T^4_{\rm eff}=-\alpha \Pi R \frac{{\rm d} \Omega}{{\rm d} R},
\end{equation}
we get an expression for the central disc temperature:
\begin{equation}\label{eq:centt}
2\sigma_{\rm SB} T^4_{\rm c}=- R\displaystyle\frac{{\rm d}\Omega}{{\rm d}R}W_{r \phi}\left[\displaystyle\frac{219}{1024}\kappa \Sigma+1\right].
\end{equation}
This expression differs from the similar equation for the standard disc (see equation 2.24 in \citealt{SS73}) because we account differently for the vertical structure of the disc.

The central pressure may be expressed as a sum of radiation and gas pressures:
\begin{equation}\label{eq:pressure}
\displaystyle\frac{15}{16}\displaystyle\frac{W_{r\phi}}{\alpha H}=\displaystyle\frac{aT_{\rm c}^4}{3}+\displaystyle\frac{3}{2}\displaystyle\frac{\Sigma k T_{\rm c}}{Hm_{\rm p}},
\end{equation}
where we have taken into account $\alpha$-prescription and the relations between the central and vertically-integrated quantities implied by the adopted vertical structure (see Appendix~\ref{sec:appendix}).

Close to the disc boundary, the infalling matter has angular velocity that differs from the angular velocity of the neutron star. 
Entering the magnetosphere, the accreting material should somehow lose the excess of angular momentum.
Following \citet{DL2014} and \citet{Aly90}, we assume that this excess angular momentum  is removed by the magnetic and radiation stresses at the magnetospheric boundary: 
\begin{equation}\label{eq:excess}
\dot M(\Omega_{\rm in}-\Omega_{\rm ns})R^2_{\rm in}=k_{\rm t}\displaystyle\frac{\mu^2 H_{\rm in}}{R^4_{\rm in}}+L\displaystyle\frac{\Omega_{\rm in}}{c^2}H_{\rm in}R_{\rm in},
\end{equation}
where $k_{\rm t}={B_{\phi}}/{B_z}$ is the ratio of magnetic field components and $L=\eta \dot M c^2$ is the luminosity released close to the neutron star, in the accretion column with efficiency $\eta$.  
 Parameter $k_{\rm t}$ can vary from $k_{\rm t}=0$ (purely poloidal magnetic field in the region near corotation radius) to $k_{\rm t}\simeq 1$ \citep{DL2014}. Some authors tend to use $k_{\rm t}=1/3$ \citep{lipunov92}. In this work we use $k_{\rm t}=0.5$.  In Section \ref{sec:impacs}, the effects of different $k_{\rm t}$ will be considered.

At the boundary, internal disc pressure is balanced with the external pressure of the radiation source $P_{\rm rad}={L}/{4\uppi R^2 c}$ and the magnetic field pressure $P_{\rm mag}={\mu^2}/{8\uppi R^{6}}$. 
Note that the radiation pressure can be even higher by a factor up to two due to reflection of photons, depending on the scattering albedo of the disc. 
Thus vertically integrated pressure at $R_{\rm in}$ is $\Pi=(P_{\rm rad}+P_{\rm mag})H_{\rm in}$. 
The $\alpha$-prescription (\ref{eq:alphapr}) gives us the vertically integrated stress at the boundary:
\begin{equation}\label{eq:wrfbound}
W^{\rm in}_{r\phi}=2\alpha H_{\rm in} \left( \displaystyle\frac{\mu^2}{8\uppi R_{\rm in}^6}+\displaystyle\frac{L}{4\uppi R_{\rm in}^2 c}\right). 
\end{equation}
 
\section{Solving the disc structure} \label{sec:solve}
\subsection{The basic system of equations}

Integrating equation (\ref{eq:momconst}),  we obtain a general expression for the vertically-integrated viscous stress:
\begin{equation} \label{eq:wrf}
W_{r\phi}=\dfrac{\dot M}{2\uppi R^2}(\Omega R^2-\Omega_{\rm in} R^2_{\rm in})+\frac{R^2_{\rm in}}{R^2}W_{r\phi}^{\rm in}.
\end{equation}
Using equations  (\ref{eq:wrf}) and (\ref{eq:alphapr}) we obtain the radial gradient of vertically-integrated pressure
\begin{equation}\label{eq:dpdr}
\begin{array}{l}
\dfrac{\uppartial \Pi}{\uppartial R}=\displaystyle\frac{1}{\alpha}\displaystyle\frac{\uppartial W_{r\phi}}{\uppartial R} 
\displaystyle = \frac{1}{\alpha}\left(\displaystyle\frac{\dot M \Omega_{\rm in} R^2_{\rm in}}{\uppi R^3}+\displaystyle\frac{\dot M}{2\uppi}\displaystyle\frac{\uppartial \Omega}{\uppartial R}-2\frac{R^2_{\rm in}}{R^3}W^{\rm in}_{r\phi}\right). \\
\end{array}
\end{equation}
Substituting expression~(\ref{eq:dpdr}) into the radial Euler equation~(\ref{eq:omega}), we have the following expression linking deviations from Kepler's law and the pressure gradient expanded from the previous equation:
\begin{equation}\label{eq:sigma}
\Omega^2 R= \displaystyle\frac{1}{\Sigma \alpha}\left(\frac{\dot M \Omega_{\rm in} R^2_{\rm in}}{\uppi R^3}+\displaystyle\frac{\dot M}{2\uppi}\displaystyle\frac{\uppartial \Omega}{\uppartial R}-2\displaystyle\frac{R^2_{\rm in}}{R^3}W^{\rm in}_{r\phi}\right) +\displaystyle\frac{GM}{R^2}.
\end{equation}
This equation is solved for $\Sigma$ at each radius. 
We still need one more equation to calculate the derivative of the rotation frequency and the unknown $\Omega(R)$.
Expressing ${\rm d}\Omega/{\rm d}R$ from equation~(\ref{eq:centt}) yields:
\begin{equation}\label{eq:domega}
\displaystyle \frac{{\rm d} \Omega}{{\rm d} R}=-\frac{T^4_{\rm c}}{R}\frac{2\sigma_{\rm SB}}{W_{r\phi}}\left(\frac{219}{1024}\kappa \Sigma+1\right)^{-1}.
\end{equation}
This is the only differential equation we need to solve. 
All the quantities in its right-hand side are found locally by solving algebraic equations: $W_{r\phi}$ is found from equation (\ref{eq:wrf}), $\Sigma$ from equation (\ref{eq:sigma}), and the central temperature can be calculated from the equation of state in the equatorial plane following from equation~(\ref{eq:pressure}): 
\begin{equation}\label{eq:temperature}
T^4_{\rm c}+\displaystyle\frac{9}{2}\displaystyle\frac{\Sigma k}{a m_{\rm p} H}T_{\rm c}-\displaystyle\frac{45}{16}\displaystyle\frac{W_{r\phi}}{\alpha a H}=0.
\end{equation}
We solve this quartic equation for $T_{\rm c}$ using Ferrari's formula, see \citet{quartic} for details.

\subsection{Dimensionless equations}

In this section we give all the equations in dimensionless form, as they were used to calculate the disc structure.
Here we list the dimensionless  parameters and combinations we use throughout the paper.  
We normalize the  neutron star mass as
\begin{equation}\label{all:mass}
m=\displaystyle\frac{M}{1.4{\rm M_{\odot}}}.
\end{equation}
The radii and disc thicknesses   $r=\displaystyle\frac{R}{R_{\rm g}}$ and $h=\displaystyle \frac{H}{R_{\rm g}}$   are measured in units of the  gravitational radius 
\begin{equation}\label{all:rg}
R_{\rm g}=\displaystyle\frac{GM}{c^2} .
\end{equation}
The angular frequency is normalized by the local  Keplerian frequency as
\begin{equation}\label{all:omega}
\omega=\displaystyle\frac{\Omega}{\sqrt{GM/R^3}}.
\end{equation}
The characteristic  magnetic moments of neutron stars lie in the range $10^{28}- 10^{32}$~G~cm$^3$, hence we normalize $\mu$ as
\begin{equation}\label{all:mu}
\mu_{30}=\displaystyle\frac{\mu}{\mu_0}=\displaystyle\frac{\mu}{10^{30}{\rm G\,cm^{3}}}.
\end{equation} 
The mass accretion rate is normalized by the Eddington value as
\begin{equation}\label{all:mdot}
\dot m=\displaystyle\frac{\dot M}{\dot M_{\rm Edd}},
\end{equation}
where
\begin{equation}\label{all:medd}
\dot M_{\rm Edd}=\displaystyle\frac{4\uppi GM}{c\kappa}\simeq2.3\times 10^{17} m \  {\rm g \, s}^{-1}.
\end{equation}
It is convenient to express the surface density in the units of the inverse opacity $\kappa^{-1}$. 
This quantity has also the physical meaning of vertical optical depth of the disc 
\begin{equation}\label{all:tau}
\tau=\kappa \Sigma.
\end{equation}
The dimensionless  version of vertically-integrated tangential stress  may be constructed as
\begin{equation}\label{all:wrf}
w_{r\phi} = \dfrac{\kappa}{c^2}W_{r\phi}.
\end{equation}
For the temperature, we use the following normalization
\begin{equation}\label{all:temp}
T_{\rm c}=t_{\rm c} T_*,
\end{equation}
where
\begin{equation}\label{all:tempstar}
T_*=\left(\displaystyle\frac{GM \dot M_{\rm Edd}}{R_{\rm g}^3 \sigma_{\rm SB}}\right)^{1/4} \!\!\!\!
=\!\! \left(\displaystyle\frac{4\uppi c^5}{\kappa GM \sigma_{\rm SB}}\right)^{1/4}\!\! \!\!
\simeq 9.6\times 10^7 m^{-1/4}\,{\rm K}.
\end{equation}
The inner radius of the disc may be normalized either by the gravitational or by the Alfv\'en radius
\begin{equation}\label{all:rin}
r_{\rm in}=\displaystyle\frac{R_{\rm in}}{R_{\rm g}}=\xi r_{\rm A},
\end{equation}
where the dimensionless Alfv\'en radius is 
\begin{equation}\label{all:ra}
r_{\rm A}=\displaystyle\frac{R_{\rm A}}{R_{\rm g}}=2^{-1/7}\left(\displaystyle\frac{\lambda \mu_{30}^2}{\dot m} \right)^{2/7} 
\end{equation}
and
\begin{equation}\label{all:lambda}
\lambda=\displaystyle\frac{\mu_0^2 c^8 \kappa}{8\uppi (GM)^5} \simeq 4\times 10^{10}m^{-5}.
\end{equation}
We also introduce  { the natural time unit} 
\begin{equation}\label{all:pstar}
p_*=\displaystyle\frac{2\uppi GM}{c^3 } \simeq 4.33\times 10^{-5}m\ {\rm s}, 
\end{equation}
and the dimensionless  factor 
\begin{equation}\label{all:chi}
\chi=\displaystyle\frac{k}{m_{\rm p}}\left(\displaystyle\frac{4\uppi}{c^3 \kappa GM \sigma_{\rm SB}}\right)^{1/4}=8.8\times 10^{-6}m^{-1/4}.
\end{equation}

From equation~(\ref{eq:wrfbound})  we can find the stress tensor at the boundary of the disc
\begin{equation}\label{eq:d.wrfbound}
w^{\rm in}_{r\phi}=\displaystyle
2 \alpha h_{\rm in}  \left(\lambda \displaystyle\frac{\mu^2_{30}}{r^6_{\rm in}}+\displaystyle\frac{\dot m \eta }{r^2_{\rm in}}\right). 
\end{equation}
The angular velocity at the inner boundary can be found from  equation~(\ref{eq:excess}): 
\begin{equation}\label{eq:d.omegain}
\omega_{\rm in}
=\displaystyle\frac{r_{\rm in}^{3/2}}{1-\eta h_{\rm in}/r_{\rm in}}\left(2\lambda \displaystyle\frac{k_{\rm t}\mu^2_{30} h_{\rm in}}{\dot m r^6_{\rm in}}+\displaystyle\frac{p_*}{p_{\rm s}} \right),
\end{equation}
where $p_{\rm s}$ is the neutron star period in seconds.
A general expression for the vertically-integrated viscous stress is then obtained from equation~(\ref{eq:wrf}):
\begin{equation}\label{eq:d.wrf}
w_{r\phi}=\displaystyle\frac{2\dot m}{r^2}(\omega \sqrt{r}-\omega_{\rm in}\sqrt{r_{\rm in}})+\left(\displaystyle\frac{r_{\rm in}}{r}\right)^2 w^{\rm in}_{r\phi}.
\end{equation}
The relative thickness of the disc follows from equation~(\ref{eq:thickness}):
\begin{equation}\label{eq:d.beta}
\beta=\displaystyle\frac{H}{R}=\sqrt{\displaystyle\frac{5 w_{r\phi} r}{\tau \alpha}}.
\end{equation}
The relative thickness of the disc at the inner boundary can be obtained from equations~(\ref{eq:d.wrfbound}) and (\ref{eq:d.beta})
\begin{equation}\label{eq:d.betain}
\beta_{\rm in}=\displaystyle\frac{10 }{\tau_{\rm in}}\left(\lambda\displaystyle\frac{\mu^2_{30}}{r^4_{\rm in}}+\dot m \eta\right). 
\end{equation}

From equations~(\ref{eq:sigma}) and (\ref{eq:domega}) we find the equation for the dimensionless angular velocity 
\begin{eqnarray}\label{eq:d.tau}
1-\omega^2&=& \displaystyle\frac{1}{\tau \alpha}\left[ -\displaystyle\frac{4 \dot m r_{\rm in}^{1/2}\omega_{\rm in}}{r} \right. \nonumber \\ 
&+&\left. \! \! \displaystyle\frac{16\uppi \dot m t^4_{\rm c} r}{m w_{r\phi}}\left(\displaystyle\frac{219}{1024}\tau+1 \right)^{-1}\! \!+ \!\displaystyle\frac{2 w^{\rm in}_{r\phi} r^2_{\rm in}}{r }\right]  ,
\end{eqnarray}
which is solved numerically for $\tau$. 
The temperature is obtained using equation~(\ref{eq:temperature}) which in dimensionless form can be written as 
\begin{equation}\label{eq:d.temperature}
\displaystyle\frac{45}{256\uppi}\displaystyle\frac{w_{r\phi}}{\alpha h}=t^4_{\rm c}+\displaystyle\frac{9\chi}{32\uppi}\displaystyle\frac{\tau t_{\rm c}}{h}.
\end{equation}
The only differential equation~(\ref{eq:domega}) in dimensionless form becomes
\begin{equation}\label{eq:d.domega}
\displaystyle\frac{{\rm d}\omega}{{\rm d}r}=\displaystyle\frac{3}{2}\displaystyle\frac{\omega}{r}-\displaystyle\frac{8\uppi t^4_{\rm c} \sqrt{r}}{w_{r\phi}}\left(\displaystyle\frac{219}{1024}\tau+1 \right)^{-1}.
\end{equation}

\subsection{Boundary conditions}

Two boundary conditions on the inner boundary of the disc are given by equations~(\ref{eq:d.wrfbound}) and (\ref{eq:d.omegain}). 
The first one arises from equation~(\ref{eq:excess}) that says that the excess of angular momentum is removed by magnetic and radiation stresses acting upon the magnetospheric boundary. 
If the neutron star and the disc are in corotation, $\Omega_{\rm ns}=\Omega_{\rm in}$, { the left-hand side } of equation~(\ref{eq:excess}) is zero, and the thickness of this disc should be zero as in the standard disc with a zero-stress boundary condition. 
If the neutron star rotates faster than the disc, the left-hand side of equation~(\ref{eq:excess}) becomes negative while the right-hand side is always positive. 
Therefore, stationary accretion becomes impossible.
This situation is known as propeller regime.  

Boundary conditions can be solved together with equations  (\ref{eq:d.tau}) and (\ref{eq:d.temperature}). 
There are 5 unknowns in this system of equations: angular velocity $\omega_{\rm in}$, magnetospheric radius $r_{\rm in}$, disc thickness $h_{\rm in}$, surface density (optical depth) $\tau_{\rm in}$ and the central disc temperature $t_{\rm in}$ for a given accretion rate, magnetic field and the spin period. 
{ This means that for any $\xi=r_{\rm in}/r_{\rm A}$, the inner radius is specified and, thus,  we can calculate  the remaining 4 parameters} at the inner boundary of the disc. 

There are two terms in the right-hand side of  equation~(\ref{eq:excess}): the first one depends on the magnetic field and accretion rate and the second one depends on the spin period of the neutron star. 
The relative importance of the two terms is set by the ratio of the neutron star period and the equilibrium period
\begin{equation}\label{eq:peq}
p_{\rm eq} \!= \! 2\uppi \left(\displaystyle\frac{\mu^2 c\kappa}{8\uppi \sqrt{2} (GM)^{8/3}\dot m } \right)^{3/7} 
\!\!\!\! \simeq 1.3\mu^{6/7}\dot m^{-3/7}m^{-8/7}\ \mbox{s} .
\end{equation}
This period is determined from the equality of the Alfv\'en and the corotation radii.\footnote{Note that this definition of the equilibrium period does not take into account factor $\xi$  in  equation (\ref{eq:alfven}) for the Alfvenic radius.} 
If the spin period of a neutron star is less than $\xi^{3/2}p_{\rm eq}$, magnetosphere rotates faster than the disc and the matter from the disc cannot accrete onto the neutron star. 
Accretion is possible only if the spin period of the neutron star is longer than equilibrium multiplied by $\xi^{3/2}$. 
If $p_{\rm s} \leq p_{\rm eq}\xi^{3/2}$, condition (\ref{eq:d.omegain}) cannot be satisfied, and the neutron star enters propeller regime. 
It is thus natural to expect the structure of the boundary conditions and the disc itself to depend primarily upon the ratio $p_{\rm s}/p_{\rm eq}$, or on the ratio $R_{\rm co} / R_{\rm A} = \left(p_{\rm s}/p_{\rm eq}\right)^{2/3}$. 
At the same time, it is interesting to consider the role of the magnetic field and of the mass accretion rate at a fixed value of $p_{\rm s}/p_{\rm eq}$.

\begin{figure*}
\includegraphics[width=\columnwidth]{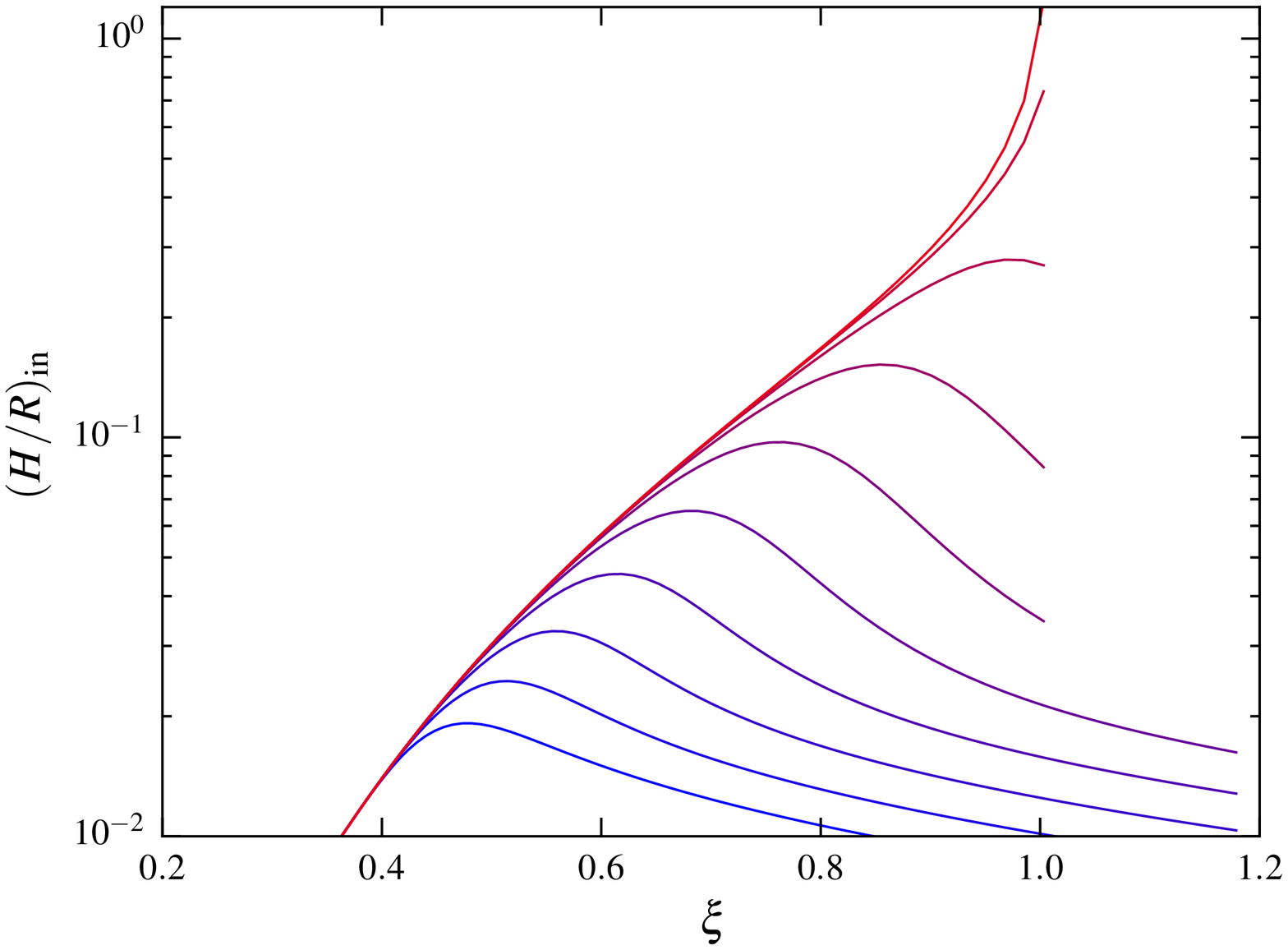}
\includegraphics[width=\columnwidth]{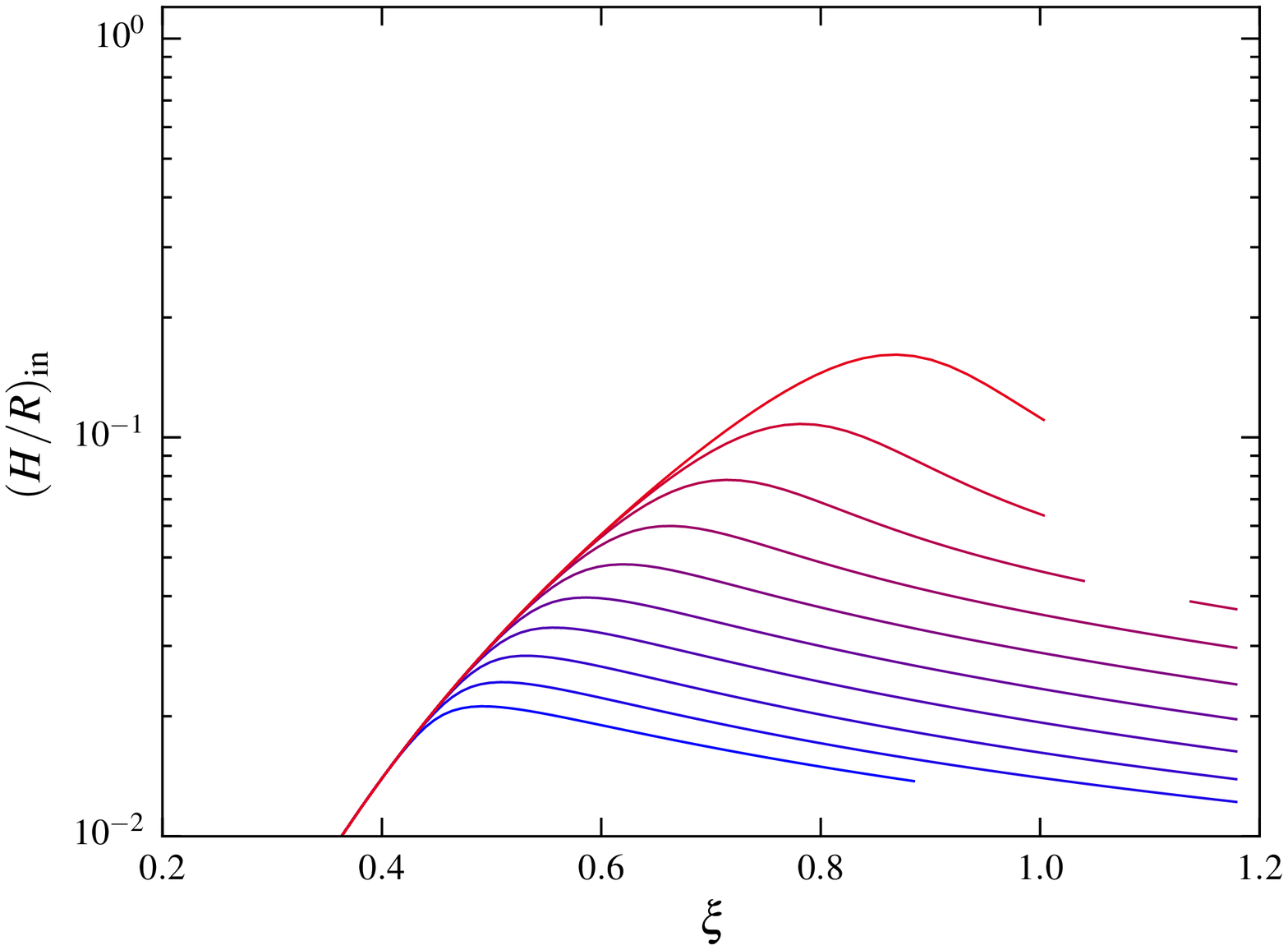}
\caption{The relative disc thickness  at the inner boundary a function of $\xi$ shown for { different accretion rates}. 
The accretion rate changes between $\dot m=0.1$ (blue curve) and $\dot m=1000$ (red curve) log-uniformly. 
A smaller accretion rate results in a smaller relative disc thickness. The spin period here is $p_{\rm s}=10 p_{\rm eq}$,  $\mu_{30}=0.1$ (left panel) and $\mu_{30}=100$ (right panel).} \label{fig:boundmdot_small} 
\end{figure*}

\begin{figure}
\includegraphics[width=\columnwidth]{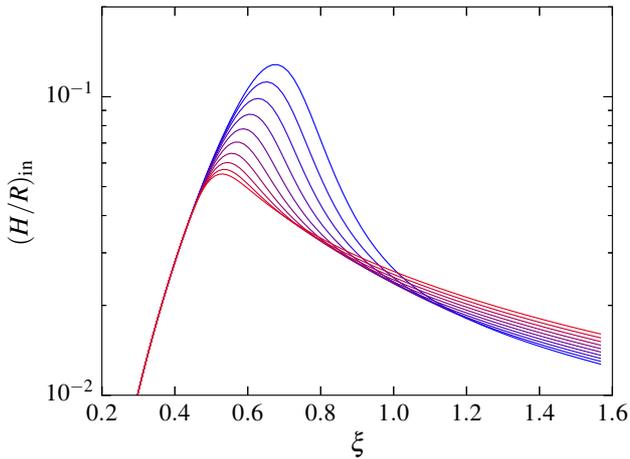}
 \caption{{ Same as Fig.~\ref{fig:boundmdot_small}, but for constant $\dot m=10$, $p_{\rm s}=10 p_{\rm eq}$  and different magnetic moments. 
The magnetic moment changes between $\mu_{30}=0.1$ (blue curve) and $\mu_{30}=100$ (red curve) log-uniformly. 
A smaller magnetic moment results in a higher relative disc thickness in the range $\xi \sim 0.5-1$.  }} \label{fig:boundmu}       
\end{figure}

\begin{figure}
\includegraphics[width=\columnwidth]{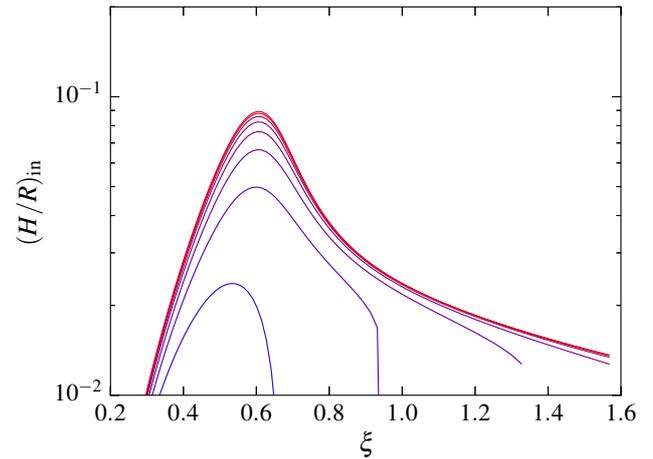}
\caption{{ Same as Fig.~\ref{fig:boundmdot_small}, but for constant $\mu_{30}=1$ and $\dot m=10$  and different spin periods. 
The spin period decreases from $p_{\rm s}=20p_{\rm eq}$  (red curve) to $p_{\rm s}=0.2p_{\rm eq}$  (blue curve). 
The maximal disc thickness increases with the spin period.  }} \label{fig:boundps} 
\end{figure}

{ The  properties of our boundary conditions are illustrated by Figs~\ref{fig:boundmdot_small}--\ref{fig:boundps}. 
In Fig.~\ref{fig:boundmdot_small}, we consider the influence of the mass accretion rate on the relative thickness at the inner disc radius. For this we fixed the magnetic moment $\mu$ and the spin period in $p_{\rm eq}$ units, $p_{\rm s}=10p_{\rm eq}$, and considered different accretion rate values. 
The value of $\xi$ is still unknown, but most of the curves have a single maximum of $(H/R)_{\rm in}$ that is the upper limit for the possible disc thickness at the inner boundary. 

The relative thickness at $r_{\rm in}$ calculated using our set of boundary conditions is plotted for different accretion rates as a function of $\xi$ in Fig.~\ref{fig:boundmdot_small} for $\mu_{30}=0.1$ (left panel) and $\mu_{30}=100$ (right panel).  
The accretion rate changes in the range $\dot m=0.1-1000$ from the blue (bottom) to the red (top) curve. 
In case of a small magnetic field, the thickness of the disc rapidly grows with the accretion rate and for high accretion rates it becomes comparable with the radius. When magnetic field is high, the disc stays thin even for high accretion rates.    

Fig.~\ref{fig:boundmu} shows similar curves for constant spin period in equilibrium units $p_{\rm s}=10p_{\rm eq}$ and accretion rate $\dot m = 10$ for different magnetic moments $\mu_{30}=0.1-100$. High magnetic fields keep the disc thin. For all the parameter values, disc thickness tends to be strictly limited at small $\xi$. }
Magnetic stress at the boundary is proportional to $r_{\rm in}^{-6}$ and increases very rapidly at small $\xi$. 
Thus, for small $\xi$, even a small thickness of the disc is sufficient to remove the excess angular momentum. 
That is why the disc becomes thin and stays very close to Keplerian rotation. 

{
Finally we study the influence of the spin period on the conditions at the disc boundary. The disc thickness at the inner disc edge is shown as a function of $\xi$ for constant magnetic moment $\mu_{30}=1$ and mass accretion rate $\dot m = 10$ and different spin periods $p_{\rm s}=(0.2-20)p_{\rm eq}$ in Fig.~\ref{fig:boundps}. Smallest spin periods produce very thin accretion discs as $\Omega_{\rm in}-\Omega_{\rm ns}$ tends to be very small. 
For large $\xi$ and small spin periods, the boundary conditions cannot be satisfied. 
 Some of the curves disappear at large $\xi$, where equation~(\ref{eq:d.tau}) does not have solutions any more. As this happens at small periods, it is logical to identify the effect with entrance into propeller regime.  An alternative explanation is that, for such parameter sets, there is no {\it stationary} solution, and the accretion disc will undergo some kind of non-stationary accretion.
}

\subsection{Searching for the solution}

The inner boundary of the disc determined by the value $\xi = R_{\rm in}/R_{\rm A}$ is, in fact, unknown but may be found by matching the boundary conditions to the solution of equation (\ref{eq:d.domega}). 
Equation (\ref{eq:d.domega}) is stable if integrated inwards.  
We find surface density from pressure gradient, hence we need to assume a small non-Keplerianity at the outer boundary of the disc. 
In our calculation we use $\omega_{\rm out}=0.99$, but this value does not influence the disc structure anywhere but in a very small region near the outer disc radius. 
The disc structure rapidly evolves toward a solution independent of the starting value of $\omega$.

We start our calculations by solving the boundary conditions.  
For a given $\xi$ we solve equations (\ref{eq:d.wrfbound}), (\ref{eq:d.omegain}),  (\ref{eq:d.tau}) and (\ref{eq:d.temperature}).  
After that we start integrating the radial structure of the disc from the outer boundary. 
For a given point we calculate viscous stresses $w_{r \phi}$ using equation (\ref{eq:d.wrf}). 
Then we substitute it to equations  (\ref{eq:d.betain}), (\ref{eq:d.tau}), and (\ref{eq:d.temperature}) and solve these equations together to find the surface density $\tau$ of the disc, disc thickness $h$ and central temperature $t_{\rm c}$.  
We substitute these quantities into equation (\ref{eq:d.domega}) and find the increment of angular velocity. 
Then we solve differential equation (\ref{eq:d.domega}) with a leapfrog integration scheme, repeating all the steps listed above in this paragraph at each radius. 

To find the disc boundary we use the shooting method: first we assume some $\xi$, using it we calculate $\omega_{\rm in}$ with the boundary condition (\ref{eq:d.omegain}) and then calculate the whole structure of the disc from the outside. 
We then vary $\xi$ until the angular velocity at the boundary is equal to $\omega_{\rm in}$ calculated from the boundary condition. 
We also require the surface density and disc thickness calculated solving the disc structure equations to coincide with the corresponding quantities at the boundary. 
Sometimes, this additional requirement allows to exclude spurious solutions.  
This intersection is found using bisection method. We search for $\xi$ in the range $0.2-2$ that usually contains one solution. The value of $\xi$ is found with accuracy $\sim 10^{-3}$. 
In some cases, for example,  for  $p\lesssim p_{\rm eq}$ or for very thick, supercritical discs (see next section) equation (\ref{eq:d.tau}) does not have any solutions. 

\section{Results}\label{sec:results}

In this Section, we consider the results of our simulations for different parameters $\mu_{30}$, $\dot m$, $p_{\rm s}$, $\eta$, $\alpha$ and $k_{\rm t}$. 
While the first three are likely to vary from one object to another and even for a single object, the other three are poorly known mainly due to unknown details of the physics involved. 
The effective value of $\eta$ is affected not only by the efficiency of accretion but also by the beaming and collimation of the radiation of the accretion column. 
The last two parameters, $\alpha$ and $k_{\rm t}$, depend on the complicated physics of magnetic field amplification and reconnection in the disc and the magnetosphere. 
Below, in Sections~\ref{sec:str} and \ref{sec:rmag}, we will assume $\eta=0$ { (i.e. no irradiation from the accretion column)}, $\alpha=0.1$, and $k_{\rm t}=0.5$. 
We will also mainly focus on the case on slowly rotating magnetosphere, that proves to be a good approximation for large mass accretion rates. 
In Section~\ref{sec:impacs}, the effects of spin period and irradiation will be considered. 

\subsection{Disc structure}\label{sec:str}

Depending on the accretion rate and the size of the magnetosphere, the disc around a neutron star can be thin and gas-pressure dominated, or it can come closer to the neutron star and be thicker and radiation-pressure-dominated in its inner parts.

\begin{figure}
  \adjincludegraphics[width=\columnwidth,trim={0.5cm 1cm 0cm 0.5cm},clip]{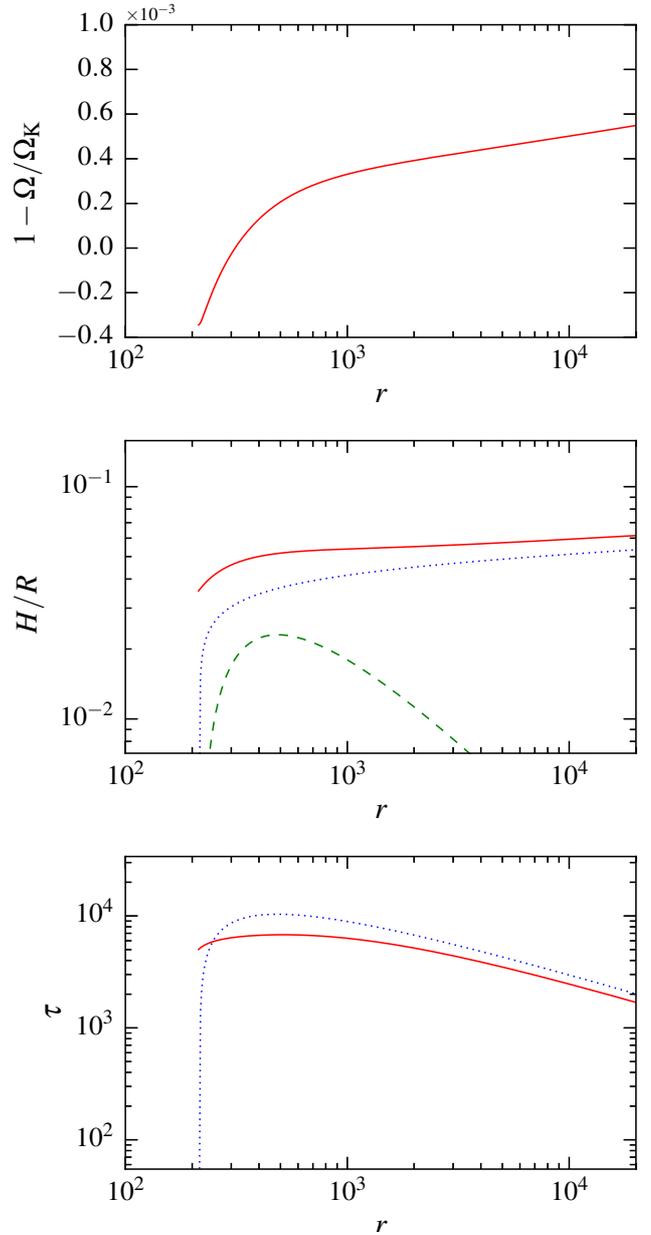} 
	\caption{Structure of the disc with  $\mu_{30}=1$, $\dot m=10$ and $p_{\rm s}=10p_{\rm eq}=4.87$ s. For these parameters $\xi=0.43$.  The upper panel shows  deviation from the Keplerian rotation $1-\Omega/\Omega_{\rm K}$ 
		as a function of the radial coordinate $r$. { Note that the deviations from Kepler's law are tiny, about $10^{-3}$.} The middle and the bottom panels show the relative disc thickness $H/R$  and the optical depth, respectively. Results for our model are shown by the solid red lines. 
The standard disc model in zones A and B is shown by the dashed green and dotted blue lines, respectively.}
\label{fig:discstr1}   
\end{figure}

The structure of a thin disc  for  $\mu_{\rm 30}=1$, $\dot m=10$ and  $p_{\rm s}=10p_{\rm eq}=4.87$ s is presented in Fig.~\ref{fig:discstr1}. 
The upper panel shows deviation from Keplerian rotation  $1-\omega$ as a function of the radial coordinate $r=R/R_{\rm g}$. 
The middle panel shows the relative thickness $H/R$ of the disc. 
Our results are shown with the solid red line. 
The thickness predicted by the standard disc model within zone A (radiation-pressure-dominated zone) or zone B (gas-pressure-dominated zone) is plotted with the dashed green or dotted blue lines, respectively. 
We use the following expressions for the disc thickness in the radiation-dominated zone:
\begin{equation}\label{eq:hssa}
H_{\rm A}=\displaystyle\frac{3\sqrt{5}}{8\uppi}\displaystyle\frac{\kappa}{c}\dot M 
\left[1-\left(\displaystyle\frac{R_{\rm in}}{R}\right)^{1/2}\right]
\end{equation} 
or in a dimensionless form: 
\begin{equation}\label{eq:hssa_d}
h_{\rm A}=\displaystyle\frac{3\sqrt{5}}{2}\displaystyle 
\dot m \left[1-\left(\displaystyle\frac{r_{\rm in}}{r}\right)^{1/2}\right].
\end{equation}
In the gas-dominated zone the disc thickness is
\begin{eqnarray}\label{eq:hssb}
H_{\rm B}& =& \sqrt{5}\left(\displaystyle\frac{9}{8}\right)^{1/10} \displaystyle\frac{1}{a^{1/10}(2\uppi)^{2/5}}\displaystyle\frac{\dot M^{1/5}}{\Omega_{\rm K}^{7/10}\alpha^{1/10}}\left(\displaystyle\frac{\kappa}{c}\right)^{1/10} \nonumber \\
\displaystyle & \times&  \left(\displaystyle\frac{k}{m_{\rm p}} \right)^{2/5}\left[1-\left(\displaystyle\frac{R_{\rm in}}{R}\right)^{1/2}\right]^{1/5} 
\end{eqnarray} 
or 
\begin{equation}  \label{eq:hssb_d}
h_{\rm B}=\sqrt{5}\,
\chi^{2/5} \displaystyle \left(\displaystyle\frac{9}{128\uppi^3}\right)^{1/10}\displaystyle\frac{\dot m^{1/5} r^{21/20}}{\alpha^{1/10}}\left[1-\left(\displaystyle\frac{r_{\rm in}}{r}\right)^{1/2}\right]^{1/5},
\end{equation}
rewriting formulas from \citet{SS73} using our dimensionless variables.
There is an additional factor $\sqrt{5}$ in both equations that accounts for the vertical structure we use. 
The correction factor $1-(r_{\rm in}/{r})^{1/2} $ arises from the boundary condition $W^{\rm in}_{r\phi}=0$ used in the standard disc model.

The bottom panel of Fig.~\ref{fig:discstr1} shows the optical depth as a function of radius. 
For the standard disc, the optical depth in the radiation-dominated zone is
\begin{equation}\label{eq:taussa}
\tau_{\rm A}=\displaystyle\frac{64\uppi}{9\alpha}\displaystyle\frac{c^2}{\kappa}\displaystyle\frac{1}{\Omega_{\rm K} \dot M \left[1-\left(\displaystyle\frac{R_{\rm in}}{R}\right)^{1/2}\right]}=\displaystyle\frac{16}{9\alpha} \displaystyle\frac{r^{3/2}}{\dot m \left[1-\left(\displaystyle\frac{r_{\rm in}}{r}\right)^{1/2}\right]},
\end{equation}
and in the gas-dominated zone:
\begin{eqnarray}\label{eq:taussb}
\tau_{\rm B}&=&\left(\displaystyle\frac{a c}{9\uppi^3 \alpha^4}\right)^{1/5}\left(\displaystyle\frac{m_{\rm p} \kappa}{k}\right)^{4/5}\dot M^{3/5}\Omega_{\rm K}^{2/5}\left[1-\left(\displaystyle\frac{R_{\rm in}}{R}\right)^{1/2}\right]^{3/5} \nonumber \\
 &=& \displaystyle\frac{4}{\chi^{4/5}}\left(\displaystyle\frac{\uppi}{9}\right)^{1/5} \dot m^{3/5} \alpha^{-4/5}r^{-3/5}\left[1-\left(\displaystyle\frac{r_{\rm in}}{r}\right)^{1/2}\right]^{3/5}.
\end{eqnarray}
Far from the inner boundary, the structure of our disc is very close to the standard gas-pressure-dominated thin disc. 
Near the inner boundary, a different boundary condition makes the disc thicker. 

\begin{figure}
  \adjincludegraphics[width=\columnwidth,trim={0.5cm 1cm 0cm 0.5cm},clip]{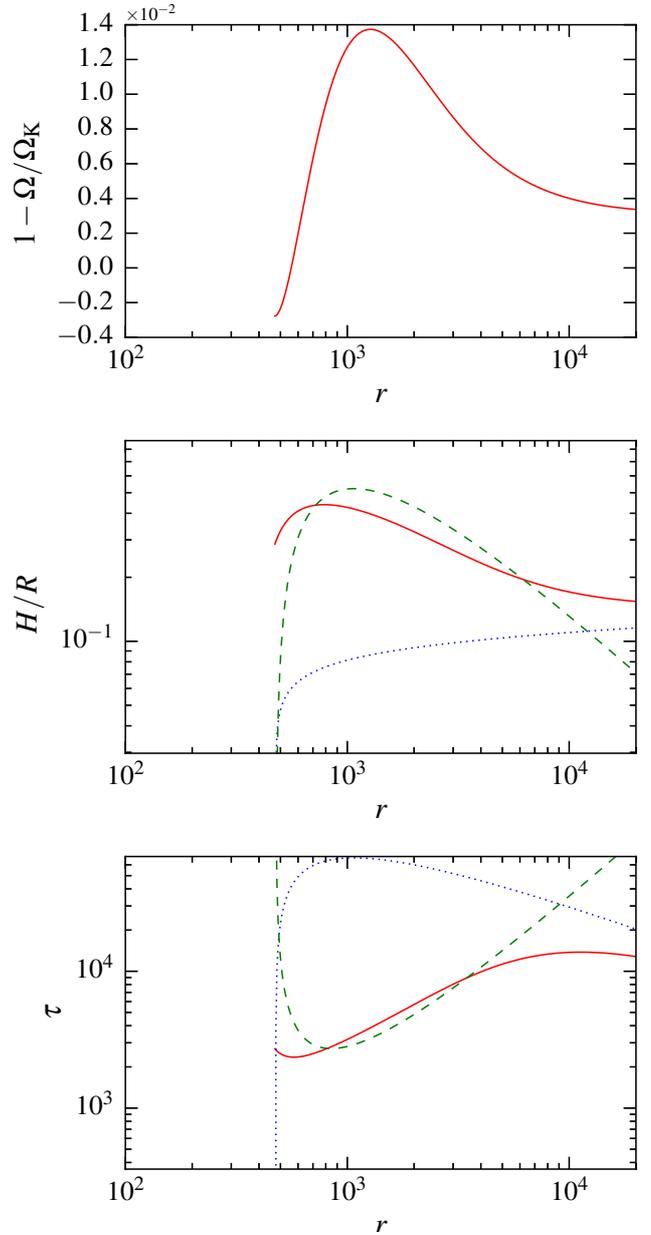} 
	\caption{Same as Fig.~\ref{fig:discstr1}, but for $\mu_{30}=10$, $\dot m=500$ and $p_{\rm s}=10p_{\rm eq}=48.65$ s, giving  $\xi=0.78$.  }
    \label{fig:discstr2}   
\end{figure}

At larger mass accretion rates and smaller magnetic moments, the inner disc becomes radiation-dominated. 
In Fig. \ref{fig:discstr2}, we show an example of a moderately-thick disc structure for $\mu_{\rm 30}=10$, $\dot m=500$ and $p_{\rm s}=10p_{\rm eq}=48.65$~s, giving  $\xi=0.78$. 
Maximal deviations from the Keplerian rotation in this case are about $1.5$ per cent, while the maximal relative thickness is about $0.5$. 
Because of the boundary conditions, the disc has a non-zero thickness at $r_{\rm in}$ but $(H/R)_{\rm max}$ is smaller than that for the standard disc because the matter is  accumulated  near the inner boundary and the disc becomes denser and thinner { because while $\Pi$ stays the same, surface density becomes higher, hence disc thickness decreases as we can see from equation~(\ref{eq:thickness})}.  
For such a disc it is important to account  possible outflows \citep{P07} together with the whole complex of effects connected to the disc thickness such as radial advection and non-local energy release \citep{AC2015}.

\begin{figure}
    \adjincludegraphics[width=\columnwidth,trim={0.5cm 1cm 0.5cm 0.5cm},clip]{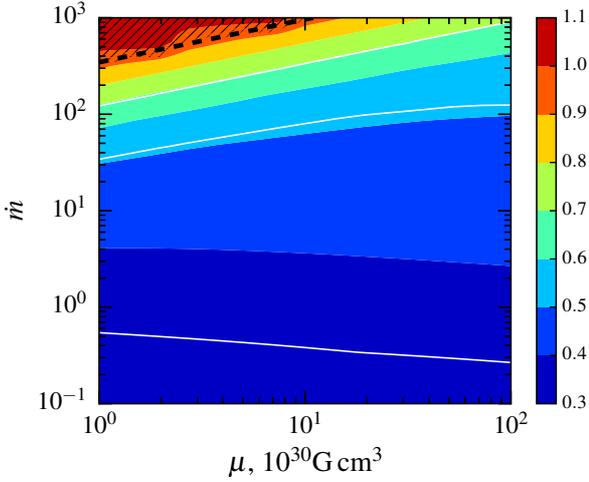}
\caption{The contours of $\xi$ shown in the $\dot m$-$\mu$ parameter space with colors. The white lines from bottom to top correspond to  $(H/R)_{\rm max}=0.03$, $0.1$ and $0.3$. Spin period is $p_{\rm s} = 10 p_{\rm eq}$. In the upper hatched region, no solution is found. The black dashed line shows the local Eddington limit, see Section \ref{sec:eddlimit} for details.  
	}\label{fig:xi}                 
\end{figure}

For the parameters characteristic for normal bright X-ray pulsars ($\mu_{30} \sim 1\div 10$, $\dot m \sim 0.1 \div 1$, see e.g. \citealt{Walter15}), the standard thin disc model is well reproduced. 
As for the standard radiation-pressure-dominated disc, the maximal $H/R$ is reached at the radius $r\simeq (1.5-2) r_{\rm in}$. 
The exact position of the maximum and its value depend on parameters $k_{\rm t}$ and $\alpha$ (see Section \ref{sec:discuss}).

Factor $\xi$ grows with the increasing disc thickness and becomes close to 1 when the relative thickness of the disc becomes large. 
As we will see below in Section~\ref{sec:rmag}, $\xi$ mainly depends on the disc thickness and the magnetic pitch parameter $k_{\rm t}$ and is relatively insensitive to other parameters. 

\subsection{Magnetospheric radius and disc thickness}
\label{sec:rmag}

The magnetospheric radius is found as the solution of the Cauchy problem that we obtain using the shooting method.  
In this section, we will consider the size of the magnetosphere and the thickness of the disc in its inner parts for different $\mu$ and $\dot m$. 
We will focus on the case of very slow rotation $p\gg p_{\rm eq}$ and $\eta=0$. 
The effects of spin period and irradiation will be considered in the next section. 

\begin{figure}
    \adjincludegraphics[width=\columnwidth,trim={0.5cm 1cm 0.5cm 0.5cm},clip]{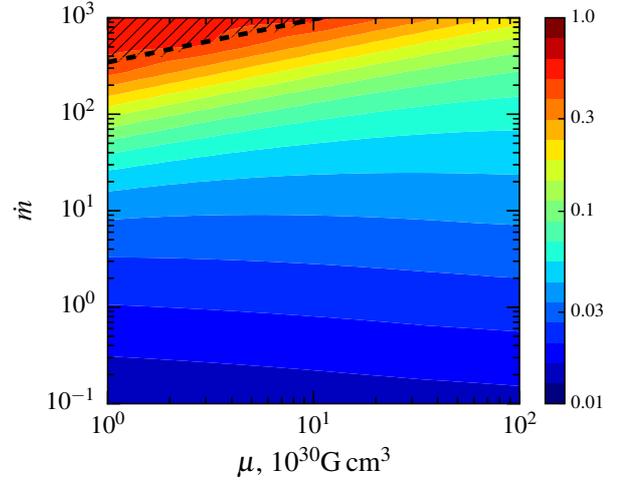} 
	\caption{The contours of $(H/R)_{\rm in}$ shown in the $\dot m$-$\mu$ parameter space with colors. The spin period scales is $p_{\rm s} = 10 p_{\rm eq}$.   }
\label{fig:hrin}
\end{figure}

Figs \ref{fig:xi} and \ref{fig:hrin}  show the contours of $\xi$  and relative thickness $(H/R)_{\rm in}$, respectively, in the $\mu_{30}$--$\dot m$ plane. 
All calculations are made for $p_{\rm s}=10p_{\rm eq}$, when the corotation radius is considerably larger than the radius of the magnetosphere. 
As long as $p_{\rm s}\gg p_{\rm eq}$, the spin period does not affect the size and the structure of the magnetosphere. 
The white lines in Fig.~\ref{fig:xi} correspond to the constant maximal relative thicknesses of the disc: $\left(H/R\right)_{\rm max}=0.03$, $0.1$ and $0.3$. 

\begin{figure}
  \adjincludegraphics[width=\columnwidth,trim={0.5cm 1cm 0.5cm 0.5cm},clip]{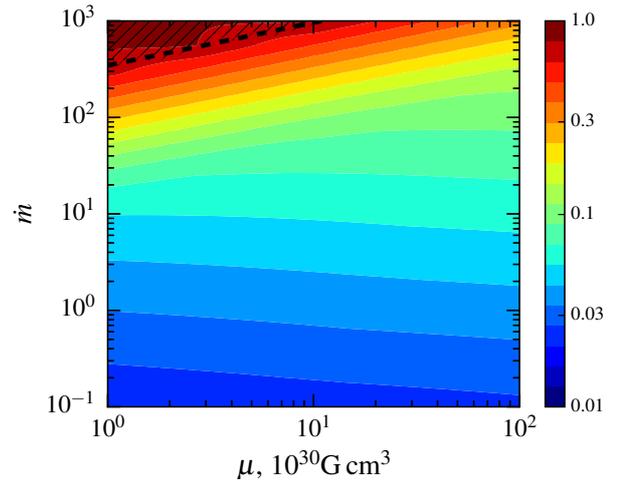} 
\caption{The contours of $(H/R)_{\rm max}$ shown in the $\dot m$-$\mu$ parameter space with colors. The spin period is  $p_{\rm s} = 10 p_{\rm eq}$.  }
\label{fig:hrmax}     
\end{figure}

As we said in the previous section, the maximal relative disc thickness  can be reached not at the boundary but at some distance of about $(1\div 2)r_{\rm in}$. 
This is similar to the case of the radiation-pressure-dominated standard disc where zero-stress boundary condition leads to a non-monotonic thickness behavior with a maximum at $\sim 2r_{\rm in}$. 
For smaller $\dot m \lesssim 10$, all the disc remains in gas-pressure-dominated state and the maximal thickness is reached at its outer rim. 
However, the value of $H/R$ in this case is practically independent of radius, except in its innermost parts (see Fig.~\ref{fig:discstr1}). 
Fig. \ref{fig:hrmax} shows the contours of $(H/R)_{\rm max}$ in $\dot m$-$\mu$ parameter space with colors. 

\begin{figure}
  \adjincludegraphics[width=\columnwidth,trim={0.5cm 1cm 0.5cm 0.5cm},clip]{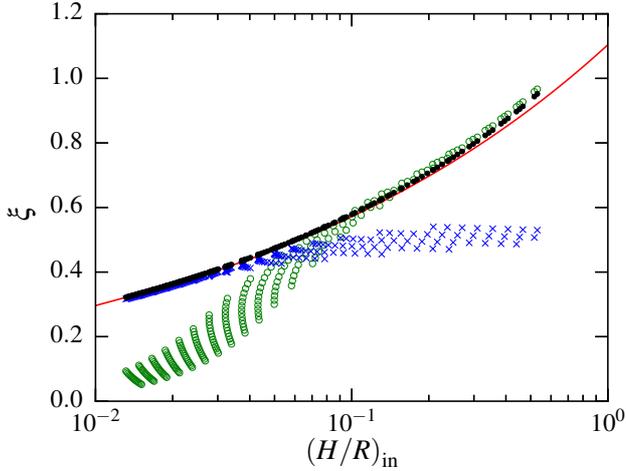}  
\caption{Dependence of disc thickness on $\xi$ (shown with black dots).  The solid red line shows the relation $\xi \propto (H/r)^{2/7}$. The green open circles and blue crosses show the slow-rotation approximations in the radiation-pressure (equation~\ref{eq:xigas}) and the gas-pressure-dominated (equation~\ref{eq:xirad}) regimes, respectively. 
}
\label{fig:xihr}                 
\end{figure}

Figs~\ref{fig:xi}--\ref{fig:hrmax} suggest that all the three quantities: $\xi$, $(H/R)_{\rm in}$ and $(H/R)_{\rm max}$, depend on each other. 
This dependence is especially tight for $\xi$ and $(H/R)_{\rm in}$.
It appears that, for any values of $\dot m $ and $\mu$, $\xi$ depends only on $(H/R)_{\rm in}$. 
This is illustrated by Fig.~\ref{fig:xihr}, showing $\xi$ as a function of $(H/R)_{\rm in}$ for the same data that were used in Figs~\ref{fig:xi} and ~\ref{fig:hrin}.
Here, the scatter around a universal dependence $\xi(H/R)$ is less than $1$ per cent. 
The results of the calculations are shown with black circles and the solid red line is the $\xi = 2^{-3/7} [ k_{\rm t} (H/R)_{\rm in}]^{2/7}$ slow-rotation asymptotic we derive below (see equation~\ref{eq:xicalc}). 

All the results discussed above were obtained for small fastness parameter $\omega_{\rm s}=\Omega_{\rm ns}/\Omega_{\rm in}\simeq (R_{\rm in}/R_{\rm c})^{3/2}$. 
This situation can be realized in transient objects during outbursts when the  accretion rate is much higher than the average. 
For such objects we can neglect the second term in the boundary condition (\ref{eq:d.omegain}). 
The boundary condition may be then solved for $\xi$ under assumption $\omega_{\rm in}=1$ that is accurate up to the terms of the order $(H/R)_{\rm in}^2$:
\begin{equation}\label{eq:xicalc}
\xi_{\infty}=2^{1/7}(2k_{\rm t})^{2/9} \left(\displaystyle\frac{\lambda \mu_{30}^2}{\dot m}\right)^{-4/63}h^{2/9}_{\rm in} ,  
\end{equation}
where $h_{\rm in}$ in general case should be  calculated numerically. 
For purely gas- and radiation-dominated standard discs this limit can be obtained analytically. 

For the gas-dominated disc it is safe to use $\xi$ in the general form given by equation~(\ref{eq:xicalc}) and the thickness follows equation~(\ref{eq:hssb}) without the correction factor, because in this case the disc thickness is almost constant everywhere. 
Expression for the magnetospheric size becomes then
\begin{equation}\label{eq:xigas}
\xi_{\infty, \rm B}=2^{1/7} \left(\displaystyle\frac{9 (2k_{\rm t})^{10} 5^5 }{128 \uppi^3}\right)^{2/69}\chi^{8/69}(\lambda \mu_{30}^2)^{2/483}  \dot m^{26/483}\alpha^{-2/69}. \end{equation}

In the radiation-pressure-dominated disc, the relative disc  thickness tends to zero more rapidly at the boundary and we cannot neglect the correction due to boundary conditions. 
The relative disc thickness  can be obtained by substituting equations~(\ref{eq:d.temperature}), (\ref{eq:d.wrfbound}) and (\ref{eq:d.betain}) to equation~(\ref{eq:d.tau}):  
\begin{equation}\label{eq:hthin}
\displaystyle\frac{h_{\rm in}}{r_{\rm in}} =\displaystyle\frac{r_{\rm in}}{\alpha} \left[\displaystyle\frac{\dot m}{r^{1/2}_{\rm in}}-\displaystyle\frac{24\dot m}{73 \alpha}\left(\displaystyle\frac{\lambda \mu_{30}^2}{r^4_{\rm in}}+\dot m \eta\right)^{-1} \right]\left(\displaystyle\frac{\lambda \mu_{30}^2}{r^4_{\rm in}}+\dot m \eta\right)^{-1}.
\end{equation}
The first factor in the right-hand side of this equation is $\sim r_{\rm A} /\alpha \gg 1$, therefore  the two terms in the square brackets should nearly cancel each other
because $(h/r)_{\rm in} \lesssim 1$. 
This gives us an estimate for $\xi$ in the radiation-pressure-dominated disc:
\begin{equation}\label{eq:xirad}
\xi_{\infty, \rm A} =2^{1/7}\left(\displaystyle\frac{73\alpha}{24}\right)^{2/9} \displaystyle\frac{\dot m^{2/7}}{(\lambda \mu_{30}^2)^{4/63}}.
\end{equation}

\begin{figure}
  \adjincludegraphics[width=\columnwidth,trim={0.5cm 1cm 0.5cm 0.5cm},clip]{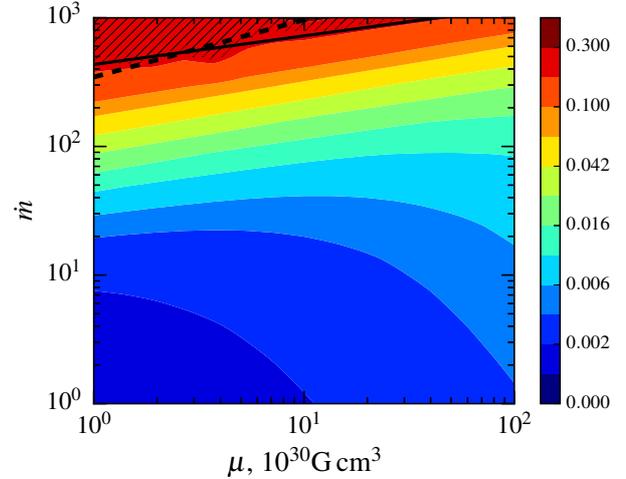}    
	\caption{The contours of difference ($\xi(\eta=0.1)-\xi(\eta=0)$) shown in the $\dot m$-$\mu$ parameter space with colors. The black dashed and solid lines  show the local and non-local Eddington limits, see Section \ref{sec:eddlimit} for details.
	}\label{fig:diff}                 
\end{figure}

The two expressions for $\xi$ given by equations (\ref{eq:xigas}) and (\ref{eq:xirad}) reproduce the scaling with $\mu_{30}$ and $\dot m$ (see Fig.~\ref{fig:xihr}) but cannot be used as universal approximations because the structure of the disc is modified near the inner boundary. 
Still they work surprisingly well for the case of a slowly-rotating magnetosphere ($p_{\rm s}\gg p_{\rm eq}$) and no irradiation ($\eta =0$). 
These two approximations are shown with green open circles and blue crosses in Fig.~\ref{fig:xihr}. 
Each point corresponds to a unique combination of $\mu_{30}$ and $\dot m$, therefore in general the abscissa and ordinate of the points are not expected to follow any single curve.
However, in their applicability regions, the two approximations show very small scatter. 

In radiation-dominated disc $\xi\propto \dot m^{2/7}$ and $r_{\rm A}\propto \dot m^{-2/7}$, therefore $r_{\rm in} = \xi r_{\rm A}$ does not depend on $\dot m$. 
Increasing accretion rate leads to an increase in the disc thickness in a way that the disc pressure remains constant. Hence, the pressure balance condition can be satisfied at the same radius for different $\dot m$. 

\subsection{Irradiation, rotation and free parameters}
\label{sec:impacs}

In Sections~\ref{sec:str} and \ref{sec:rmag} we ignored the effects of irradiation by setting $\eta=0$.  
However, the accretion efficiency of a neutron star can be $\eta=0.1-0.2$. 
Radiation of the central X-ray source affects the pressure balance condition (\ref{eq:d.wrfbound}) as well as, through radiation drag, the other boundary condition (\ref{eq:d.omegain}). 
As a result of these two effects, the inner radius becomes larger, especially for thicker discs, and the disc thickness at the inner boundary also increases. 
The difference in $\xi$ for $\eta=0$ and $\eta=0.1$ is shown in Fig.~\ref{fig:diff}. 
Thick discs are more sensitive to accretion efficiency because the intercepted luminosity fraction is proportional to the solid angle subtended by the disc $\propto ( H/R)_{\rm in}$.  
As expected, the radiation does not affect the magnetospheric radius in the thin gas-dominated disc regime, but for a thick disc, radiation can increase magnetospheric radius by about $30$ per cent. 
Irradiation increases the inner relative thickness while $(H/R)_{\rm max}$ stays about the same for a disc with and without irradiation. 

\begin{figure}
  \adjincludegraphics[width=\columnwidth,trim={0.3cm 0.8cm 0.5cm 0.5cm},clip]{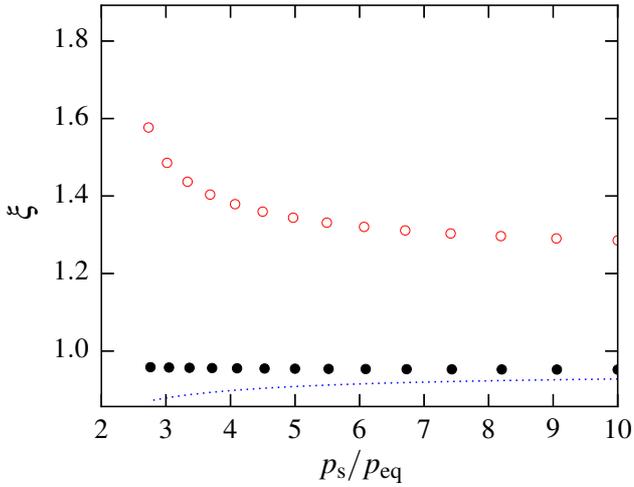}
\caption{Dependence of $\xi$ on the relative neutron star period $p/p_{\rm eq}$.  Black filled circles are plotted for the case without irradiation ($\eta=0$), and the red open circles are for $\eta=0.1$. The blue dotted line shows calculated $\xi_{\infty}$ for actual $(H/R)_{\rm in}$ (according to equation \ref{eq:xicalc}).  Here we assumed a very high acceretion rate of $\dot m=1000$ and $\mu_{30}=10$.}\label{fig:both_eff}  
\end{figure}

\begin{figure*}
  \adjincludegraphics[width=\columnwidth,trim={0.3cm 0.8cm 0.5cm 0.5cm},clip]{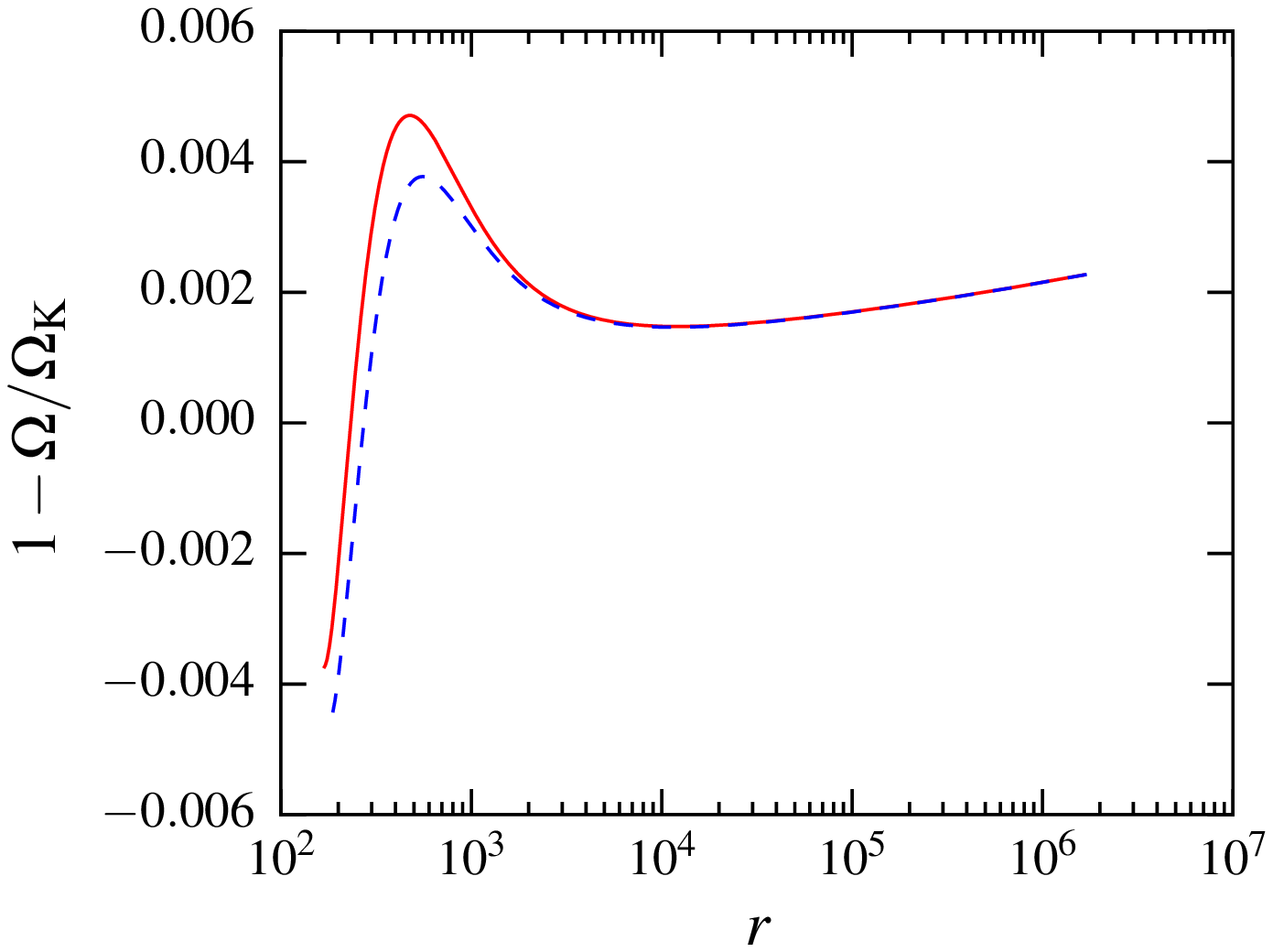}
  \adjincludegraphics[width=\columnwidth,trim={0.3cm 0.8cm 0.5cm 0.5cm},clip]{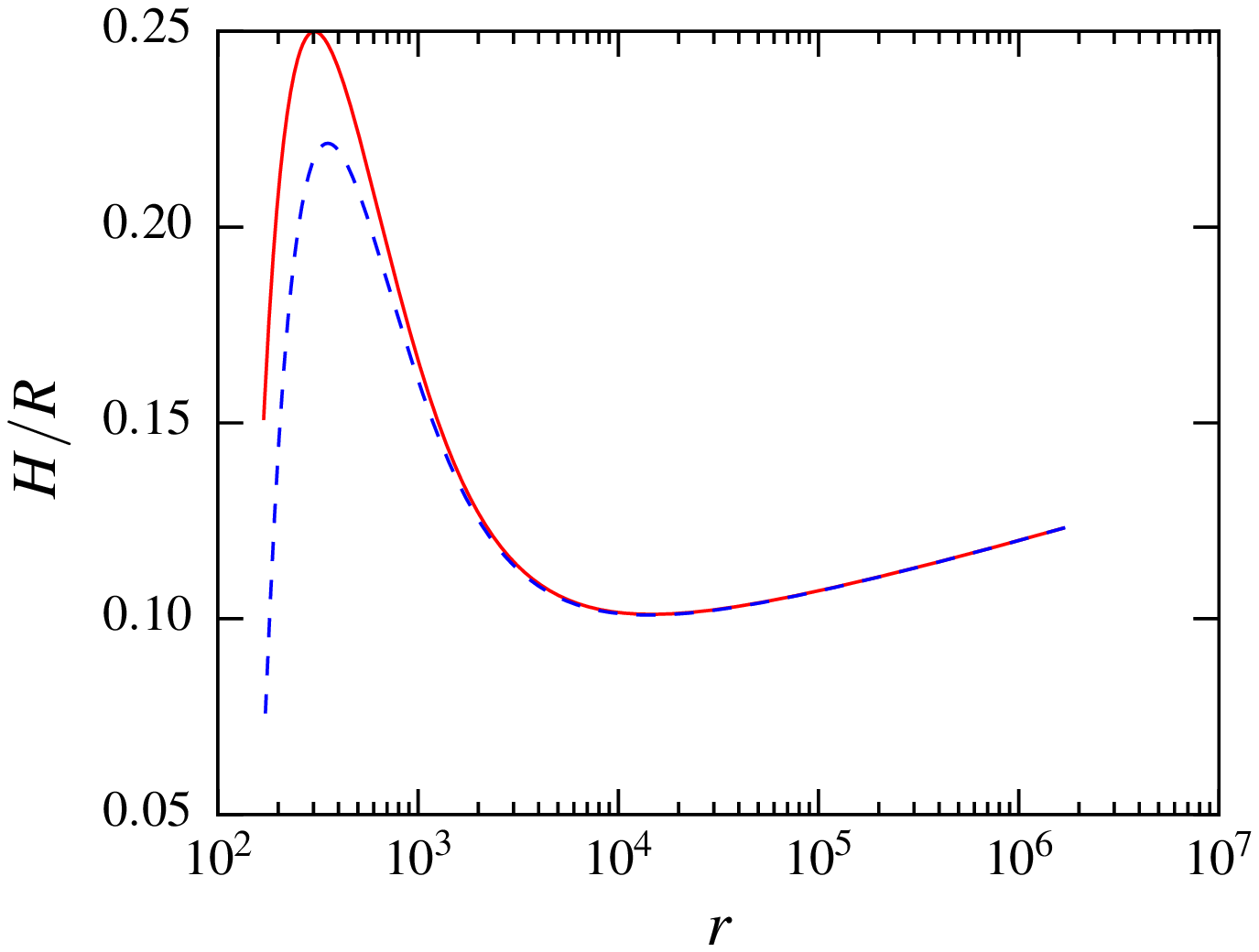}
\caption{Dimensionless angular velocity $1-\omega$ (left panel) and the relative disc thickness $(H/R)$ (right panel) as functions of $r$.  The results for the case close to corotation, $p_{\rm s}=p_{\rm eq}$, is shown by the blue dotted lines and  those for  $p_{\rm s}=10p_{\rm eq}$
by the red solid lines. Parameters are $\mu=1$ and $\dot m=100$.  }\label{fig:corotstr}                 
\end{figure*}

Though the slow-rotation estimate~(\ref{eq:xicalc}) was derived under the assumption $\eta=0$, it holds to a good accuracy (of about $\eta (H/R)_{\rm in}$) even when both $\xi$ and the disc thickness are strongly affected by irradiation.
In Fig.~\ref{fig:both_eff}, we show the effects of irradiation and spin period of the neutron star accreting at a very high rate ($\dot m = 10^3$, $\mu_{30}=10$).  
The dependence of $\xi$ on the relative neutron star spin period  $p_{\rm s}/p_{\rm eq}$ is shown with and without irradiation for a large mass accretion rate, when the structure of the disc is especially vulnerable to irradiation. 
The black circles show $\xi$ for disc without irradiation ($\eta=0$), and the red open circles are for the disc irradiated from inside ($\eta=0.1$). 
The blue dotted line shows $\xi_{\infty}$ calculated according to equation ~(\ref{eq:xicalc}). 
The slow-rotation approximation given by expression~(\ref{eq:xicalc}) works well whenever the period is much larger than the equilibrium and there is no irradiation, and gives the lowest possible value of $\xi$ for given $\mu$ and $\dot m$. 
In the radiation-dominated regime with $\eta=0$, the physical size of the magnetosphere practically does not depend on the spin period of neutron star, as it does not depend on the mass accretion rate.

In general case, in the fast-rotation limit, near corotation, disc becomes thinner, and $\xi$ increases. In Fig.~\ref{fig:corotstr}, we show how variations in neutron star spin frequency affects the structure of the disc. 
Solutions exist above the poorly constrained minimal period that we interpret as the propeller limit and discuss in more detail in Section~\ref{sec:propeller}.

\begin{figure*}
\includegraphics[width=0.7\textwidth]{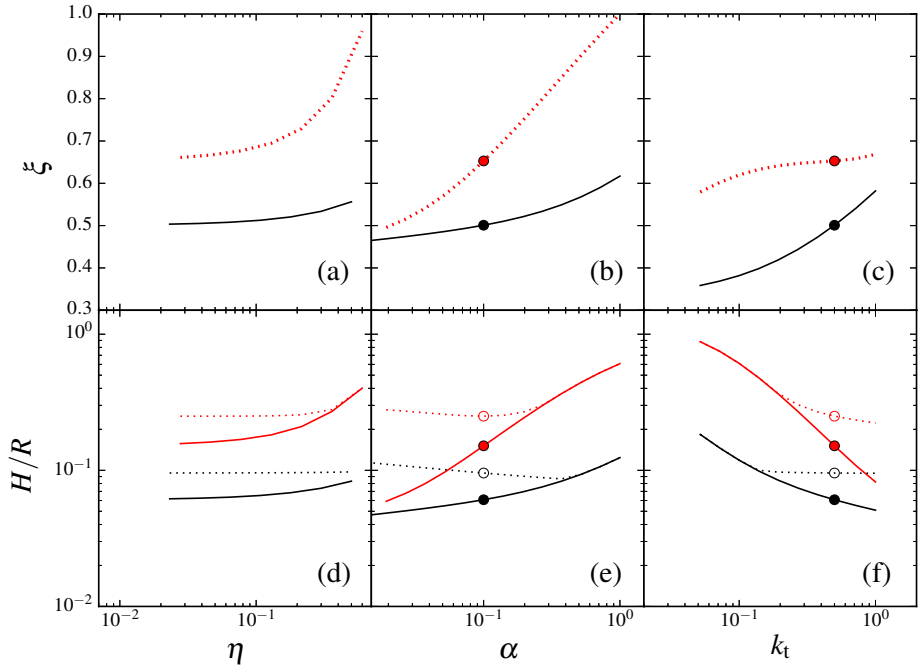}
\caption{The relative magnetosphere size $\xi$ (upper panels) and the relative disc thickness (lower panels) as functions of three free parameters of our model: the radiative efficiency of the central source $\eta$ (left), the viscosity parameter $\alpha$ (middle), and the magnetic field deformation parameter $k_{\rm t}$. The black and red curves correspond to $\mu_{30}=100$ and $\mu_{30}=1$, respectively. The mass accretion rate here is $\dot m =100$. 
In the lower panels, $(H/R)_{\rm in}$ is shown by the solid and $(H/R)_{\rm max}$ by the dotted lines. The basic models with $\alpha=0.1$ and $k_{\rm t}=0.5$ are shown by circles, red for $\mu_{30}=1$ (relatively thick disc model) and black for $\mu_{30}=100$ (thinner disc case). The open circles in the lower panels correspond to the maximal disc thickness. 
}\label{fig:all_effects}        
\end{figure*}

There are two additional free parameters in the model: viscosity of a disc $\alpha$ and the ratio of magnetic field components $k_{\rm t} = B_\phi / B_z$. 
Their ratio $\alpha/k_{\rm t}$ has an important impact on accretion disc structure, being the ratio of tangential components of the viscous stress inside the disc $W_{r\phi}=2\alpha H_{\rm in}\displaystyle\frac{\mu^2}{8\uppi R^6_{\rm in}}$ to the stress created by the magnetic fields in the magnetosphere $W_{r\phi}=2k_{\rm t} H_{\rm in}\displaystyle\frac{\mu^2}{4 \uppi R^6_{\rm in}}$.
In the limiting case when these stresses are equal, $\alpha=2 k_{\rm t}$, position of the relative thickness $(H/R)_{\rm max}$ maximum in the disc reaches the inner boundary.  
Effects of variations in $\eta$, $\alpha$ and $k_{\rm t}$ are shown in Fig.~\ref{fig:all_effects} for two different magnetic moments. 
As it is easy to see from this figure, the maximal disc thickness is insensitive to $\eta$, a large $\alpha / k_{\rm t} \gtrsim 1$ make the inner disc thicker (maximal thickness is reached at $r_{\rm in}$), and larger viscosity generally makes the magnetospheric size  larger and the disc thicker. 

Though the radiative efficiency of neutron star accretion seems to lie somewhere around $\eta \simeq 0.15 -0.2$, irradiation is also affected by anisotropy of the radiation emitted by the accretion column, that can add a factor of two uncertainty in the effective value of $\eta$. 
Besides, if some of the accreting material is expelled from the magnetosphere (that is likely if the radiation pressure is large, see Section~\ref{sec:eddlimit}), the mass accretion rate in the accretion column becomes smaller, this proportionally decreases the effective value of $\eta$. 
This justifies a large range of possible efficiencies considered in Fig.~\ref{fig:all_effects}.

\subsection{Deviations from Kepler's law}

{ All the disc structures discussed above show very small deviations from the Keplerian rotation because the discs are quite thin.  
Non-Keplerianity increases with disc thickness as $1-\omega \sim (H/R)^2$. According to the boundary condition equation~(\ref{eq:excess}), the disc thickness at the inner boundary increases when parameter $k_{\rm t}$ becomes smaller. 
For small $k_{\rm t}\lesssim \alpha/2$, the disc thickness strongly peaks at the inner boundary (see Fig.~\ref{fig:all_effects}), and non-Keplerianity can become significant.  

	\begin{figure}
		\adjincludegraphics[width=\columnwidth]{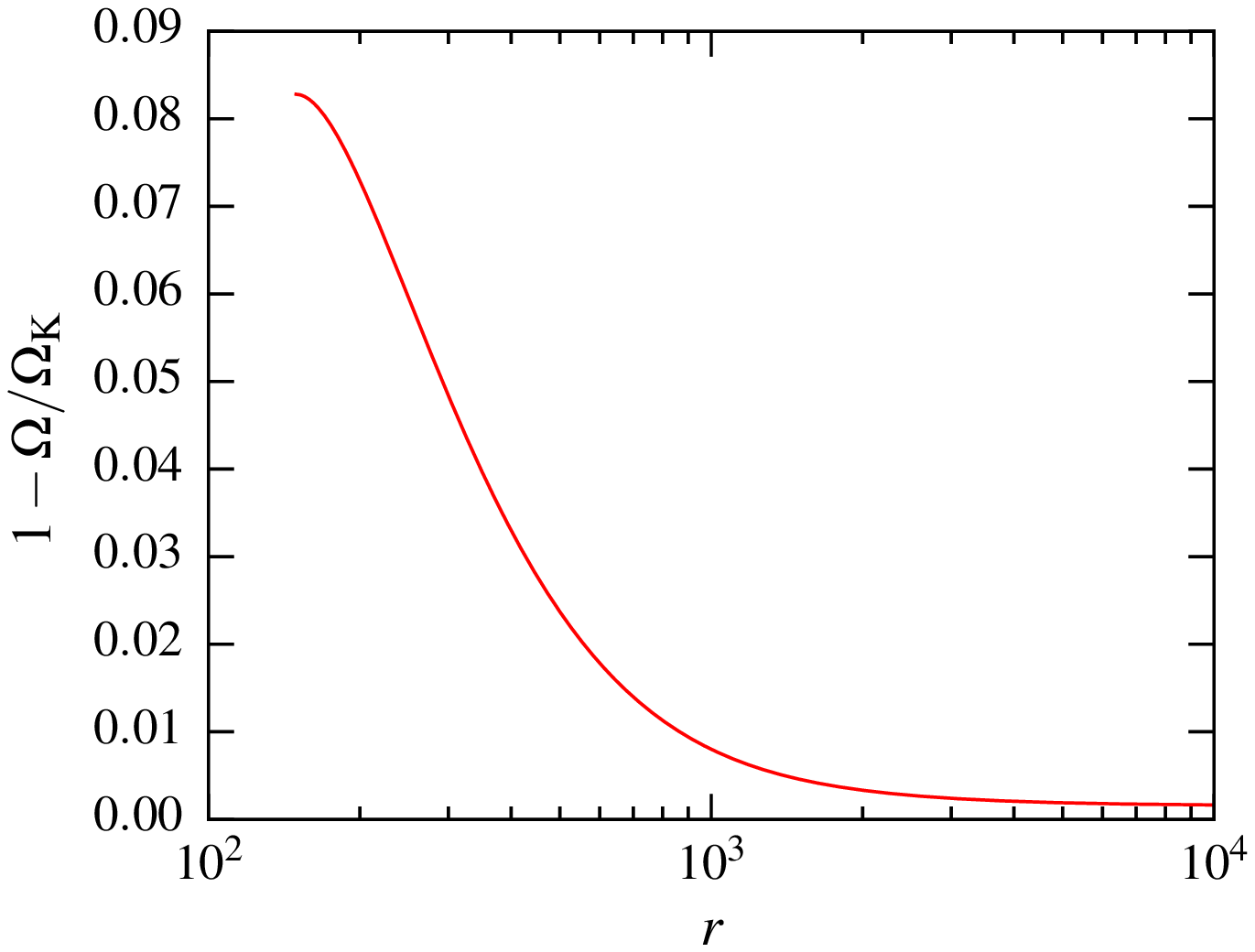} 
		\caption{The  deviation from the Keplerian rotation $1-\Omega/\Omega_{\rm K}$ as a function of the radial coordinate $r$ is plotted for the disc with  $\mu_{30}=1$, $\dot m=100$ and $p_{\rm s}=10p_{\rm eq}=0.9$s, $k_{\rm t}=0.05$. For these parameters $\xi=0.58$.  } 
		\label{fig:smallkt}   
	\end{figure}
	
	The structure of a thin disc with $k_{\rm t}=0.05$ for a neutron star with $\mu_{\rm 30}=1$, $\dot m=100$ and  $p_{\rm s}=10p_{\rm eq}=0.9$~s is presented in Fig.~\ref{fig:smallkt}. 
	The difference from Keplerian rotation  $1-\omega$ is shown as a function of the radial coordinate $r$. Here non-Keplerianity can reach $\sim 10$ per cent and the thickness $(H/R)_{\rm in}\simeq1$. 
For even thicker discs, deviations from Kepler's law can formally reach even larger values, but the model itself becomes inapplicable. 
	
	 }

\section{Discussion} \label{sec:discuss}

\subsection{Magnetosphere in radiation-dominated regime}

A neutron star accreting at a high enough rate will have the relative magnetosphere size given by equation~(\ref{eq:xirad}) that implies the physical size of magnetosphere
\begin{equation}\label{eq:magnetospheric_radius}
r_{\rm in} = \xi_{\infty, {\rm A}} r_{\rm A} = \left(\displaystyle\frac{73\alpha}{24}\right)^{2/9}(\lambda \mu_{30}^2)^{2/9}
\end{equation}
independent of the mass accretion rate. 

The rate of changes of the total angular momentum of a NS can be written down as follows:
\begin{equation}\label{eq:torques}
I\displaystyle\frac{{\rm d}\Omega_{\rm ns}}{{\rm d}t}=N_{\rm up}+N_{\rm down},
\end{equation}
where  $N_{\rm up}=\dot M\sqrt{GMR_{\rm in}}$ is the spin-up torque due to accretion, $N_{\rm down}$ is the spin-down torque. 
If accretion rate is large enough we can neglect the second term.  
In the radiation-pressure-dominated disc,  $R_{\rm in}$ does not depend on the accretion rate and $|\dot p_{\rm s}|\propto \dot M$.  
For the gas-pressure-dominated disc, $|\dot p_{\rm s}|\propto \dot M^{427/483}$ ($427/483 \simeq 6/7$ with the accuracy of about 3 per cent). 
These are the two limiting cases, and for real objects $|\dot p_{\rm s}| \propto \dot M^{6/7} - \dot M$, that is in a good agreement with observations of X-ray pulsars \citep{Bildsten97}. 

\subsection{Eddington limit(s)}\label{sec:eddlimit}

The challenge in understanding the super-Eddington accretion is in the large radiation fields that seemingly prevent the accretion that causes them. 
In the case of the accretion discs of ULX-pulsars, one should take into account two radiation fields: one from the disc itself and one from the central X-ray source (accretion column) emitting most of the luminosity. 
Correspondingly, we will talk about local and non-local Eddington limits. 
Impact of the second, non-local radiation field becomes important when its pressure becomes comparable to the magnetic field pressure 
\begin{equation}\label{eq:radfrac}
\displaystyle 
\frac{P_{\rm rad}}{P_{\rm mag}} = \left. \frac{L}{4\uppi R_{\rm in}^2 c} \middle/ \frac{\mu^2}{8\uppi R_{\rm in}^6} \right. = 2^{-4/7} \eta \left(\frac{\lambda \mu_{30}^2}{\dot m}\right)^{1/7} \xi^4.
\end{equation}
If we now substitute expression~(\ref{eq:xirad}) for $\xi$, the pressure ratio becomes
\begin{equation}\label{eq:radfrac1}
\displaystyle 
\frac{P_{\rm rad}}{P_{\rm mag}} \! = \! \left( \frac{73\alpha}{24}\right)^{8/9}\!\!\!\!\! \eta \dot m \left( \lambda \mu_{30}^2\right)^{-1/9} \simeq 0.5 \frac{\eta}{0.1} \left( \frac{\alpha}{0.1}\right)^{8/9}\!\!\! \mu_{30}^{-2/9} \dot m .
\end{equation}
For large mass accretion rates, and also large magnetic field and viscosity, irradiation from accretion column is important. 
Usually, the adopted picture of magnetospheric accretion assumes that matter is captured by magnetic field at the boundary of the disc. 
If the radiation pressure from the accretion column is larger than the magnetic pressure, optically thin infalling matter may be blown away by the radiation pressure force  escaping the magnetosphere \citep{M17}. 
Optically thick accretion flow, on the other hand, can remain bound, but its dynamics will still be  affected by the radiation pressure. 

Let us estimate the critical (non-local Eddington) mass accretion rate when magnetic pressure equals radiation pressure. 
When the two pressures are equal, equation~(\ref{eq:radfrac1}) implies:
\begin{equation}\label{eq:dotmcrit}
\dot m_1=\displaystyle\frac{1}{\eta} \left(\displaystyle\frac{24}{73\alpha}\right)^{8/9} (\lambda \mu^2_{30})^{1/9}\simeq 430 \frac{0.1}{\eta} \left( \frac{0.1}{\alpha}\right)^{8/9} \mu_{30}^{2/9}.
\end{equation}
When accretion rate is larger than this value, our model is incomplete as some part of the accreting material is probably ejected from the magnetosphere. 
This limitation is plotted on Fig.~\ref{fig:diff} by the black solid line. 

The local Eddington limit is reached when $(h/r)_{\rm max}=1$. 
In our model relative thickness is maximal at $r\simeq 2r_{\rm in}$. 
Let us consider a standard radiation-dominated disc far from the corotation. 
Using the disc thickness from equation ~(\ref{eq:hssa}) and estimating $\xi$ from equation (\ref{eq:xirad}), one can obtain the maximum possible accretion rate:
\begin{eqnarray}
\dot m_{2} & \simeq &  \displaystyle\frac{4}{3\sqrt{5}}\left(1-\displaystyle\frac{1}{\sqrt{2}}\right)^{-1}(\lambda \mu_{30}^2)^{2/9}\left(\displaystyle\frac{73\alpha}{24}\right)^{2/9}  \nonumber \\ 
& \simeq &  350 \left( \frac{\alpha}{0.1}\right)^{2/9} \mu_{30}^{4/9}.
\end{eqnarray}
This limit is plotted in Figs~\ref{fig:xi}--\ref{fig:hrmax} and \ref{fig:diff} by the black dashed lines. 
For moderate magnetic moments $\mu_{30}\sim 1\div 10$, both limits are comparable $\dot m_1 \sim \dot m_2 \sim 500$, but for larger magnetic fields there is a region in parameter space where only non-local Eddington limit is violated. 
Unlike the local limit, $\dot m_1$ depends strongly on viscosity and practically does not depend on the magnetic field strength. 

\subsection{Propeller regime and ULX}\label{sec:propeller}

The first ULX-pulsar, M82~X-2 \citep{Bach14}, has a high enough luminosity of $L_{\rm X}\sim 10^{40}$ erg s$^{-1}$ for its inner parts of the disc to be radiation-pressure-dominated. 
\citet{Tsygankov16} assumed that M82 X-2 is very close to corotation { (see also  \citealt{Bach14}, \citealt{Lyutikov14}, \citealt{DO15} for more discussion on this topic)}, and we observe it in two states: the low luminosity corresponds to the propeller regime and the high luminosity to the accretion regime. 
Therefore, the observed period of $p_{\rm s}=1.37$~s is the minimal possible period of a neutron star with a given magnetic field when accretion is still possible. 

In the propeller regime, the angular velocity at the inner boundary of the disc is equal to the angular velocity of the neutron star. 
In this case, condition~(\ref{eq:excess}) implies $h_{\rm in}=0$. 
In the radiation-pressure-dominated regime, it leads to expression~(\ref{eq:hthin}) that is approximately valid whenever the disc is radiation-pressure-dominated and irradiation is negligible. 
Usually, the size of the magnetosphere is not known directly, but we may assume $\omega_{\rm in}=1$ and thus $p_{\rm s} = p_* r_{\rm in}^{3/2}$.
This allows to estimate the magnetic field as a function of the (accurately measured) spin period and (less constrained) mass accretion rate and efficiency
\begin{equation}\label{eq:muest}
\mu_{30}=\displaystyle\frac{(p_{\rm s}/p_*)^{4/3}}{\lambda^{1/2}}\left(\displaystyle\frac{24 (p_{\rm s}/p_*)^{1/3}}{73\alpha}-\dot m \eta \right)^{1/2}.
\end{equation}
If the accretion column radiates isotropically, $\eta \dot m = L/L_{\rm Edd}$. 
In a more general anisotropic case, $\eta$ is not bound to be equal to the apparent radiative efficiency of neutron star accretion. 
Isotropic efficiency is well approximated by 
\begin{equation}
\eta \simeq \frac{GM}{R_*c^2} \simeq 0.2 m \frac{10\,{\rm km}}{R}.
\end{equation}
At some luminosity, expression in the brackets  in equation (\ref{eq:muest}) becomes zero. 
This luminosity coincides with the non-local Eddington limit~(\ref{eq:dotmcrit}) if the disc is in co-rotation $p_{\rm s} = p_* r_{\rm in}^{3/2}$. 
For case of M82~X-2 ($p_{\rm s} = 1.37$~s, $\dot m =500$ corresponding to the isotropic luminosity $L=10^{40}$~erg\,s$^{-1}$), $\alpha=0.1$ and $k_{\rm t}=0.5$, equation~(\ref{eq:muest}) yields $\mu\simeq 5.1\times 10^{31}$G cm$^3$ if irradiation is turned off. 
The magnetospheric radius in this case corresponds to $\xi\simeq 0.7$.
Setting $\eta=0.1$ decreases the required magnetic field to $\mu\simeq 3.7\times 10^{31}$~G~cm$^3$ and $\xi \simeq 0.8$. 
These magnetic moment values correspond, respectively, to magnetic fields at the pole of $B=10^{14}$~G and $B=7.4\times 10^{13}$~G for the neutron star radius of $R_{*}=10$~km. 
The magnetic field value estimated without irradiation is in a good agreement with  that of \citet{Tsygankov16}. 
As we see, irradiation from the accretion column decreases magnetic field by a factor of about 1.4, if a standard set of auxiliary parameters is used. 
However, the limit~(\ref{eq:muest}) has strong dependence on viscosity, and may be used to get an upper limit for $\alpha$. 
It gives $\alpha \lesssim 0.21$ for efficiency $\eta=0.1$ and $\alpha \lesssim 0.1$ for $\eta=0.2$. 
The estimated magnetic moments for different $\eta$ and $\alpha$ are presented in Table~\ref{tab:m82x2}.

\begin{table}
\begin{center}
  \begin{tabular}{ | l | r | r | r | r | r | r |  }
    \hline
   \backslashbox{$\eta$}{$\alpha$}   & $0.01$ & $0.05$ & $0.1$ & $0.5$ & $1$ \\ \hline
     $0$& 161 & 72  & 51 & 22 & 16  \\
     $0.1$ & 157  & 62 & 36 & -- & --  \\ 
     $0.2$ & 153 & 52 & 10 & -- & --   \\
    \hline
  \end{tabular}
  \caption{Magnetic moments $\mu_{30}$ for different values of $\eta$ and $\alpha$. 
  All the calculations were made for $\dot m=500$ ($L=10^{40}$~erg ~s$^{-1}$) and $p_{\rm s}=1.37$~s, aimed to reproduce the properties of ULX-pulsar M82~X-2.} \label{tab:m82x2}
\end{center}
\end{table}

Another ULX-pulsar, NGC\,7793~P13 \citep{Israel17}, has a period of $p_{\rm s}=0.42$~s and a luminosity of $L_{\rm X}=5\times10^{39}$~erg~s$^{-1}$. 
Assuming that the observed period is the minimal period the magnetic field is $\mu_{30}=7\times 10^{30}$~G~cm$^3$ and $\xi \simeq 0.7$ in the case without irradiation and $\mu_{30}=8.8\times 10^{30}$~G~cm$^3$ and $\xi \simeq 0.7$ for the case with irradiation. 
The maximal relative thickness of the disk is $(H/R)_{\rm max}=0.2$, the disc is nearly standard. 
We see that for this object irradiation is not so important. 

The third ULX-pulsar in NGC\,5907 \citep{2017Sci...355.817I} has a very high apparent luminosity of $L_{\rm X}= 1.5\times10^{40}-10^{41}$~erg~s$^{-1}$ and its period was changing very rapidly from $p_{1}=1.43$~s to $p_{2}=1.13$~s during past 11 years. 
From equation (\ref{eq:dotmcrit}) we can obtain the minimal magnetic field that can explain such a large luminosity, $\mu=6\times 10^{34}$~G~cm$^3$ which is rather unphysical. 
The most probable cause for this discrepancy is the unusually large luminosity of this object, probably affected by geometrical beaming. 
A beaming factor of several can reduce the luminosity and leads to results similar to that for M82~X-2. 
Still, for such objects with extreme mass accretion rates, the disc may become super-critical, and we need to consider effects like outflows and advection to explain the properties of the object. 

Let us estimate period derivative $\dot p_{\rm s}$ expected  for the three ULX-pulsars. 
Neglecting the unknown spin-down torque component in  equation ~(\ref{eq:torques}) provides an upper limit for the period derivative. 
For the radiation-dominated disc, the magnetospheric radius is given by equation ~(\ref{eq:magnetospheric_radius}), that gives us the period derivative:
\begin{eqnarray}\label{eq:pdot}
\left|\dot p_{\rm{spin-up}}\right|  & = &  \displaystyle\frac{2(GM)^2}{I c^2 \kappa}
\left(\displaystyle\frac{73\alpha}{24}\lambda \mu^2_{30}\right)^{1/9}p^2_{\rm s}\dot m \nonumber \\
 & \simeq& \displaystyle 3\times 10^{-12}I_{45}^{-1}\dot m \mu^{2/9}_{30}p^2_{\rm s}\ {\rm s\, s^{-1}}, 
\end{eqnarray}
where $I=I_{45} 10^{45}$g\,cm$^{2}$ is the moment of inertia of a neutron star. 
For M82 X-2 we have $|\dot p_{\rm{spin-up}}|=(6.3-6.7)\times 10^{-9}$~s\,s$^{-1}$. 
The observed period derivative (see observation 007 in Table 3 of \citealt{Bach14}) is $|\dot p_{\rm obs}|=8\times 10^{-12}$~s\,s$^{-1}$ that is much smaller than the expected spin-up rate. 
There are episodes of spin-up and spin-down mentioned in the paper, so we can conclude that the object is close to the  equilibrium (see also \citealt{Eksi15}, \citealt{Lyutikov14} and  \citealt{DO15}).
The most interesting ULX-pulsar in NGC~5907 has the maximal observational spin-up rate of $|\dot p_{\rm obs}|=9.6\times 10^{-9}$~s\,s$^{-1}$ \citep{2017Sci...355.817I} which is two times smaller than we obtain from equation~(\ref{eq:pdot}) $|\dot p_{\rm s}|=2\times 10^{-8}$~s\,s$^{-1}$. 
If we trust the luminosity estimate, this object should be in strong spin-up.

\subsection{The role of advection}\label{sec:advection}
{
In our model we assume that the energy released in the disc is radiated locally from its surface. However, radiation transfer may be altered by radial advection if  the mass accretion rate is high enough.  Advection is important when the timescale for radial motion becomes comparable to or smaller than the timescale of vertical radiation diffusion.
The characteristic timescale for radial drift is:
\begin{equation}
t_{R}\sim \displaystyle\frac{R}{v_{R}},
\end{equation}
where 
\begin{equation}
v_R = \frac{\dot M}{4\pi R \Sigma}
\end{equation}
is the radial velocity.
The diffusion time scale is:
\begin{equation}
t_{\rm diff}\sim \displaystyle\frac{H^2}{5D},
\end{equation}
where the factor of 5 accounts for the fact that the energy release is distributed in height, and the effective optical depth is smaller by the same factor (see  optical depth multiplier in equation~\ref{eq:centt}). 
Substituting expression for $D$ from equation~(\ref{eq:dif}) we get the ratio of the two timescales:
\begin{equation}
\displaystyle\frac{t_{\rm diff}}{t_{R}}\simeq \displaystyle\frac{3\dot m h}{5r^2}\simeq \displaystyle\frac{2}{5}\displaystyle\frac{r_{\rm sp}}{r}\displaystyle\frac{h}{r},
\end{equation}
where $r_{\rm sp} \simeq \displaystyle\frac{3}{2}\dot m$ \citep{SS73} is the spherization radius, inside which the standard disc model is expected to break down due to  large disc thickness. 
As long as the local Eddington limit is not reached in the disc, or, equivalently, the radius of the magnetosphere is $r_{\rm in} \gtrsim r_{\rm sp}$ and $h_{\rm in}\lesssim r_{\rm in}$, accretion disc remains sub-critical, and the effects of advection, as well as other effects related to disc thickness, remain of minor importance. 
We will consider all these effects, together with the effects of outflows, in a separate paper. }


\section{Conclusions}

We have developed a new model of the accretion disc around a strongly magnetized neutron star accreting at a high, possibly super-Eddington, rate. 
Calculating the radial structure of the disc allowed us to find more accurately the position of the disc-magnetosphere interface. 
Further outside, the disc very slightly deviates from the standard model. 
The correction factor to the classical Alfvenic radius is of the order unity, but becomes larger with increasing disc thickness reaching values  of about two.  

At high mass accretion rates and high magnetic moments, the pressure of the radiation field produced by the central source (accretion column) becomes stronger than the magnetic field pressure at the magnetospheric boundary meaning that the dynamics of the inner parts of the disc becomes strongly affected by the irradiation from the inner source. 
For magnetic moments $\mu \gtrsim 10^{30\div 32}$~G~cm$^3$ (position of this boundary strongly depends on accretion efficiency and viscosity parameter), this effect becomes important well below the accretion rate when the local Eddington limit is reached in the accretion disc. 

The properties of ULX-pulsars can be generally explained in our model with magnetic moments of the order $10^{31}$~G~cm$^3$. 
The magnetar-scale fields proposed by \citet{Tsygankov16} for M82 X-2 may be over-estimated by a factor of $1.5-5$ for $\eta=0.1-0.2$ because the large disc thickness and the irradiation by the central source increase the size of the magnetosphere. 
However, the corrections due to irradiation cannot be very high. 
To surpass the Eddington limit in the accretion column and allow accretion at $\dot m \gtrsim 500$, the local magnetic field should be of the order $\sim 10^{14}$~G \citep{MST15}. 

We also showed that for large enough disc thicknesses of $H/R \gtrsim 0.1$, when the inner disc becomes radiation-pressure-dominated, the size of the magnetosphere does not depend on the mass accretion rate. 
This fact should be reflected in the variability properties of the objects with high enough accretion rates ($\dot m \gtrsim 10$) and should also affect the spin-up rate dependence on the luminosity. 
However, the existing observational data do not allow yet to check these predictions. 

\section*{Acknowledgements}

This research was supported by the Center for International Mobility,  the V\"ais\"al\"a Foundation, Nordita Visiting PhD Fellowship (AC), the Foundations' Professor Pool, the Finnish Cultural Foundation,  the National Science Foundation grant PHY-1125915 (JP) and the Academy of Finland grant 268740 (PA).  
AC and PA acknowledge support from RSF grant 14-12-00146 (vertical disc structure).
The authors are grateful to Phil Armitage, Andrei Beloborodov, Wlodek Kluzniak, Dong Lai, Galina Lipunova, Nikolai Shakura, Sasha Tchekhovskoy and Sergey Tsygankov for useful discussions.

\bibliographystyle{mnras}
\bibliography{mybib}

\appendix
\section{Vertical disc structure}\label{sec:appendix}

The vertical structure of a thin accretion disc depends on the opacity and pressure sources and is normally calculated by solving simultaneously the equations of vertical hydrostatic balance and radiation release and transfer. 
We cannot completely ignore the vertical structure because transition between vertically-integrated quantities like $\Sigma$ and equatorial-plane quantities like $\rho_{\rm c}$ requires some assumptions about the dependence of local quantities on $z$. 
To avoid calculating vertical structure for every point, we adopt a unified density dependence on relative vertical coordinate 
\begin{equation}\label{density_gen}
\rho = \rho_{\rm c} \left( 1-x^2\right)^s,
\end{equation}
where $x=z/H$. 
The hydrostatic equilibrium equation
\begin{equation}
\displaystyle \frac{{\rm d}P}{{\rm d}z} = -\rho \frac{GM}{R^3}z
\end{equation}
allows to calculate the vertical pressure profile knowing that of the density as
\begin{equation}\label{pressure_gen}
P = P_{\rm c} \left( 1-x^2\right)^{s+1},
\end{equation}
where $s\simeq 1$ is some coefficient. 
According to \citet{SSZ78}, $s=1.17$ for the radiation-dominated disc with convection. 
We will hereafter use $s=1$ unless otherwise stated. 
For $s=1$, central pressure and density are related as
\begin{equation}
P_{\rm c} = \rho_{\rm c} \dfrac{GM}{R^3} \dfrac{H^2}{4}.
\end{equation}
Thus, vertically integrated pressure and surface density are:
\begin{equation}
\Pi=\int P {\rm d}z=P_{\rm c}H \int (1-x^2)^2 {\rm d}z 
= \dfrac{16}{15}P_{\rm c} H, 
\end{equation}
\begin{equation}
\Sigma=\int \rho {\rm d}z=\rho_{\rm c} H \int (1-x^2) {\rm d}z 
= \dfrac{4}{3}\rho_{\rm c} H. 
\end{equation}

Now let us show the derivation of relation (\ref{eq:surft}) between the surface temperature and the central temperature in the disc.
The local energy dissipation rate:
\begin{equation}
\displaystyle \frac{{\rm d}F}{{\rm d}z}=\alpha P R\left| \displaystyle\frac{{\rm d}\Omega}{{\rm d}R}\right|.
\end{equation}
 Integrating this expression along vertical direction from 0 to $z$ and using equation (\ref{pressure_gen})  we get the  energy flux:
\begin{equation}
 F=\alpha P_{\rm c} RH\displaystyle \left|\frac{{\rm d}\Omega}{{\rm d}R} \right|\left[\displaystyle\frac{z}{H}-\displaystyle\frac{2}{3}\left(\displaystyle\frac{z}{H}\right)^3+\displaystyle\frac{1}{5}\left(\displaystyle\frac{z}{H}\right)^5 \right].
 \end{equation}
Then we substitute this flux into the vertical diffusion equation $F=-D\nabla_{z} \epsilon=-\displaystyle\frac{c}{3\kappa \rho} {\rm d}(aT^4)/{\rm d}z$, taking $\rho$ from equation (\ref{density_gen})  we get:
\begin{equation}
T_{\rm s}^4=T^4_{\rm c}-\displaystyle\frac{\alpha P_{\rm c} RH^2}{ac} \left|\frac{{\rm d}\Omega}{{\rm d}R}\right| 3\kappa \rho_{\rm c} \displaystyle\frac{73}{360},
\end{equation}
where $T_{\rm s}$ and $T_{\rm c}$ are the surface and central temperatures, respectively. 
Assuming that the surface temperature equals the effective temperature 
\begin{equation}
\displaystyle 2\sigma_{\rm SB} T^4_{\rm eff}=\alpha \Pi R \left|\frac{{\rm d} \Omega}{{\rm d} R} \right|,
\end{equation}
we get an expression for the central disc temperature:
\begin{equation}\label{E:ts}
2\sigma_{\rm SB} T^4_{\rm c}=- R\displaystyle\frac{{\rm d}\Omega}{{\rm d}R}W_{r \phi}\left[\displaystyle\frac{219}{1024}\kappa \Sigma+1\right].
\end{equation}

{ In the paper we were mainly interested in the objects with high luminosities. The inner parts of their discs are in radiation-pressure-dominated regime, that justifies the usage of $s=1$. However, if we consider gas-pressure-dominated disc, $s$ can be larger than 1, about $2-3$ \citep{KS98}. For $s=2$ and $3$ the thickness of the disc is expressed as
\begin{equation}\label{eq:thickness2}
H(s=2)=\displaystyle\sqrt{\frac{7\Pi}{2\Sigma}\frac{R^3}{GM}}
\end{equation}
and
\begin{equation}\label{eq:thickness3}
H(s=3)=\displaystyle\sqrt{\frac{3\Pi}{\Sigma}\frac{R^3}{GM}},
\end{equation}
respectively. 
Equation~(\ref{E:ts}) derived for different values of $s$ from 1 to 3 has a very stable form, with the expression in the brackets always equal to $1+\left(0.21\pm 0.01\right)\kappa\Sigma$. 
Energy release in the disc is very much the same for any vertical structure. 
The resulting $\xi$ for different vertical structures differs by no more than $10$ per cent, mainly as a result of varying disc thickness multiplier (compare equations \ref{eq:thickness}, \ref{eq:thickness2} and \ref{eq:thickness3}), that changes by less than $25$ per cent. 

}

\label{lastpage}

\end{document}